\documentclass[journal=apchd5,manuscript=article]{achemso}
\setkeys{acs}{doi=true}
\usepackage{amsmath}
\usepackage{xcolor}

\usepackage{graphicx}
\graphicspath{ {Images/} }

\usepackage{xr}

\makeatletter
\newcommand*{\addFileDependency}[1]{
  \typeout{(#1)}
  \@addtofilelist{#1}
  \IfFileExists{#1}{}{\typeout{No file #1.}}
}
\makeatother

\newcommand*{\myexternaldocument}[1]{%
    \externaldocument{#1}%
    \addFileDependency{#1.tex}%
    \addFileDependency{#1.aux}%
}

\myexternaldocument{SupInf}

\usepackage[version=3]{mhchem} 
\usepackage[labelfont=bf,textfont=md]{caption} 
\usepackage{caption}
\author{Jos\'e Luis Monta\~{n}o-Priede}
\affiliation[CFM]{Centro de Física de Materiales (CFM-MPC), CSIC-UPV/EHU, Paseo de Manuel Lardizabal 5, 20018 Donostia - San Sebastián, Spain}
\email{joseluis.montano@ehu.eus}
\author{Marek Grzelczak}
\affiliation[CFM]{Centro de Física de Materiales (CFM-MPC), CSIC-UPV/EHU, Paseo de Manuel Lardizabal 5, 20018 Donostia - San Sebastián, Spain}
\alsoaffiliation[DIPC]{Donostia International Physics Center (DIPC), Paseo de Manuel Lardizabal 4, 20018 Donostia - San Sebastián, Spain}
\email{marek.g@csic.es}

\title{Thermoplasmonics under optically coupled regime: A Numerical Study of Dimers, Nanolenses, and Switchable Clusters}


\begin{document}

\begin{tocentry}
\centering
  \includegraphics[height=4.5cm]{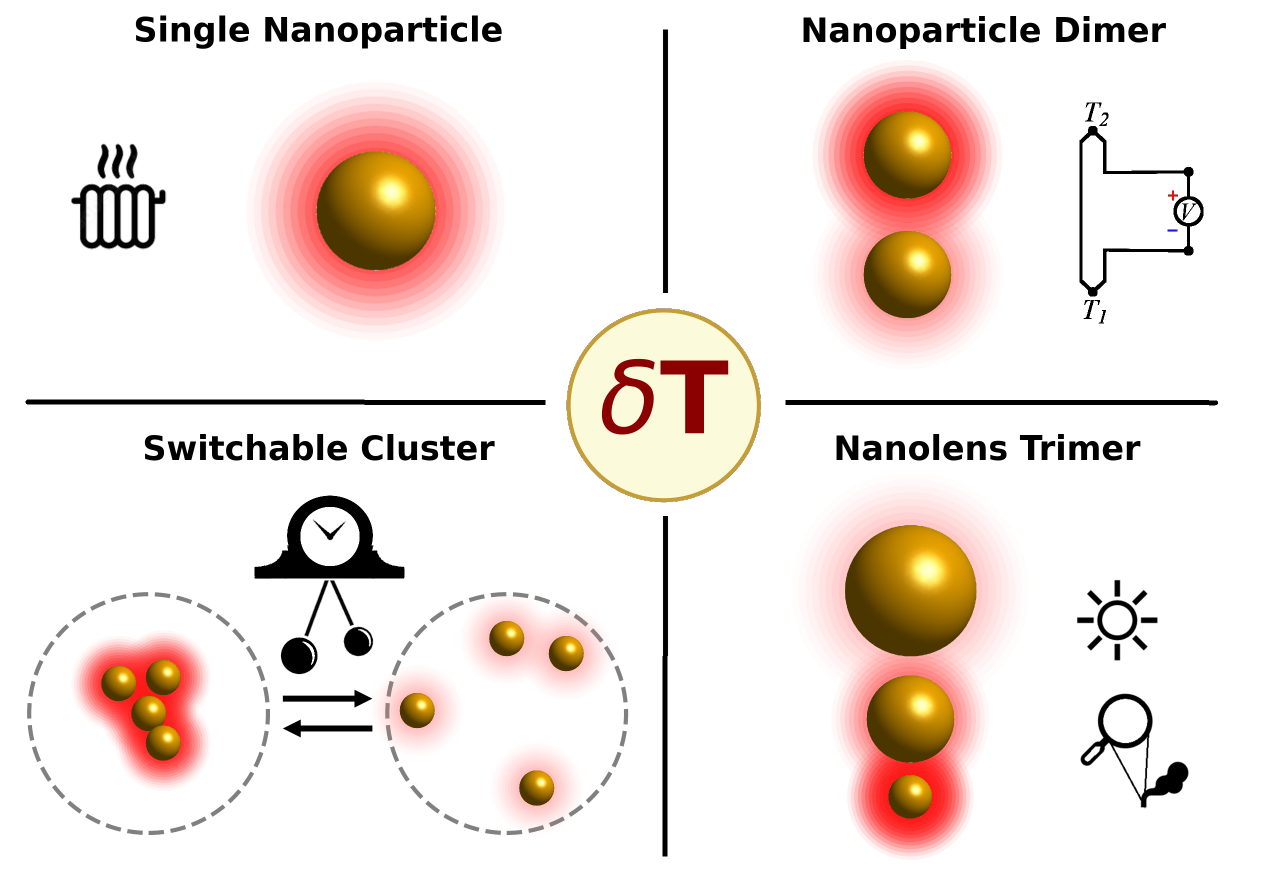}
\end{tocentry}


\begin{abstract}
The management of thermal effects in plasmonic nanostructures is frequently viewed as a detrimental waste rather than a useful, controllable entity. We show that optical coupling of plasmonic nanoparticles enables precise spatiotemporal control over nanoscale heating. Through numerical investigation of experimentally-achievable systems from individual nanoparticles and dimers to nanolenses and switchable clusters, we demonstrate how plasmon hybridization and near-field coupling dictate the magnitude and spatial distribution of temperature. Our results highlight the critical role of polarization and gap distance in tuning the thermal output of dimers, the ability of a trimer nanolens to focus heat into a sub-diffraction volume, and the pronounced thermal difference in a switchable nanoparticle cluster. This work establishes a framework for designing advanced thermoplasmonic systems where heat is not merely a detriment, but a dynamically controllable element for applications in catalysis, health, or active photonic devices.
\end{abstract}

\section{Introduction}

Light-matter interaction in plasmonic nanomaterials\cite{Yu2019,JLMP2016} gives rise to phenomena including surface-enhanced Raman spectroscopy \cite{Marzan2020,Kant2024}, enhanced photoluminescence\cite{Dahal2021, MOR2025}, or colorimetric / refractive-index sensing \cite{Li2015, Kant2024}. These phenomena experience drastic boost under optically coupled regime, a condition of small interparticle distances rendering the so-called plasmon hybridization, thereby producing intensely localized electric fields not achievable with isolated particles \cite{Nordlander2004,Halas2011}. 
Over last decades, much effort has been put into the precise geometrical engineering of such coupled nanostructures to tailor the optical response, controlling not only the enhancement of the field but also its directionality within the structures \cite{Anger2006, Hugall2018, Montano2024}.
On the other hand, due to the lossy character of plasmonic nanoparticles, the heat dissipation by the decay of plasmons via non-radiative pathways is an intrinsic and inevitable feature\cite{Baffou_2017}, being also the foundation of thermoplasmonics. The light to heat conversion has been successfully harnessed for biomedical applications \cite{Kim2019, Baffou2020} and catalysis \cite{Yang2022,Verma2024}, in which the design goal is to maximize the total heat output.\cite{GOVOROV2007, Baffou2010, Baffou2013, Herzog2014, Bell2015, Jiang2019, Adhikari2020, Fahad2024, Yu2025} The next frontier in thermoplasmonics, however, moves beyond mere heat generation towards the precise spatiotemporal control of temperature at the nanoscale \cite{Baffou2013,Baffou2020}.
For instance, Chen \textit{et al}.\cite{Chen2020} engineered gold sphere-rod nanodimers to spectrally match the solar irradiance spectrum for efficient solar thermal conversion, demonstrating control over the spectral absorption. Furthermore, Manjavacas \textit{et al}.\cite{Zundel2023} showed that long-range plasmonic coupling in bipartite nanoparticle array can be tailored to selectively heat one nanoparticle within the lattice unit cell, enabling controlled nanoscale thermal localization. 
Finally, Baffou \textit{et al}. \cite{Baffou2010} demonstrated that in a chain of closely spaced gold nanoparticles, plasmon coupling enables precise control over the temperature profile, allowing the position of the hottest spot to be shifted along the chain by tuning the illumination wavelength to selectively excite localized or propagating plasmon modes. These works underscore that optical coupling between nanoparticles is an effective design principle for managing thermal gradients.

Here, inspired by the experimentally-achievable complex and discrete plasmonic constructs, we explore numerically the spatial distribution of heat at the nanoscale. We begin with the fundamental building blocks: a single nanoparticle and a dimer, to establish the baseline size-, gap-, and polarization-dependence of the thermoplasmonic response. We then advance to a nanolens trimer, a structure distinguished for its ability to focus electric fields at small volumes \cite{Li2003,Lloyd2017}. 
We demonstrate that it similarly acts as a nano-heat-lens, capable of concentrating an intense temperature rise within a highly localized volume. This confirms the potential of such structures for applications requiring spatially confined thermal fields. 
Finally, we extend our analysis to a reversibly clustering plasmonic nanoparticles. While such system has been demonstrated experimentally for optical switching \cite{Snch2018}, we numerically evaluate the corresponding change in the thermoplasmonic response between aggregated and dispersed states. This dual-state behavior results in a large, reversible difference in the thermal output of the system upon optical excitation at a fixed wavelength.
Our work provides a coherent framework for understanding and designing coupled plasmonic systems where heat generation is controllable, directional, and integral to the system's function. This represents a shift from treating heat as a parasitic effect to be minimized, toward leveraging it as a central design element.


\section{Methods}


The thermoplasmonic response of the studied nanoparticle systems was modeled using a Green's function approach based on the Laplace Matrix Inversion (LMI) method developed by Baffou \textit{et al} \cite{Baffou2010_GreenF}. In this work, we systematize the LMI approach into a clear, sequential computational procedure, making it more accessible for implementation and extension to multi-particle systems. This structured framework enables the efficient computation of steady-state temperature distributions in complex plasmonic nanostructures under optical illumination, from the electromagnetic response to the final thermal map.

\subsection{Electromagnetic response calculation}

The optical response of metallic nanoparticles under continuous plane-wave illumination was computed using the boundary element method as implemented in the MNPBEM toolbox \cite{hohenester2012mnpbem}. This approach solves Maxwell's equations for metallic nanoparticles embedded in dielectric environments by discretizing particle surfaces into boundary elements and computing surface charge and current distributions. The dielectric function of gold and silver was modeled using experimental data from Johnson and Christy \cite{Johnson1972}. The surrounding medium was water with dielectric constant of $\epsilon_{m}=1.777$. The simulations were conducted in two stages. First, the extinction, scattering, and absorption cross-section spectra were computed in the wavelength range of 500 nm to 1100 nm for gold nanoparticles and from 300 nm to 600 nm for silver nanoparticles.

Then, for each nanoparticle system, the electric field enhancement distribution $|\mathbf{E}(\mathbf{r})/\mathbf{E}_0|^2$ was calculated at the absorption maximum wavelength, $\lambda_{\text{abs}}$, where $|\mathbf{E}_0|^2=2I_0/\epsilon_0n_mc$ is the incident field amplitude, $I_0 = 10^9$ W/m$^2$ is the incident irradiance, $\epsilon_0$ is the vacuum permittivity, and $c$ is the speed of light in vacuum. For this, the simulation domain was discretized using $N$ cubic cells of volume $\Delta V=\Delta x\Delta y\Delta z$ each.

\subsection{Heat generation calculation}

A subset of $M$ computational cells (with $M\subset N$) distributed across the P nanoparticles forming the systems was selected. For each cell $i\in M$ located at position $\mathbf{r}_i$, the heat power density is calculated as\cite{Baffou2013}:

\begin{equation}
q(\mathbf{r}_i) = \frac{\omega I_0}{n_m c} \text{Im}[\epsilon(\mathbf{r}_i)] \left|\frac{\mathbf{E}(\mathbf{r}_i)}{\mathbf{E}_0}\right|^2,
\label{eq:heat_source_density}
\end{equation}
where $\omega = 2\pi c/\lambda_{\text{abs}}$ is the angular frequency, $n_m = \sqrt{\epsilon_m}$ is the refractive index of the surrounding medium, and $\text{Im}[\epsilon(\mathbf{r}_i)]$ is the imaginary part of the dielectric function at cell $i$.

The total heat power generated in nanoparticle $k\in P$ is:

\begin{equation}
Q_k = \sum_{i \in \mathcal{M}_k} q(\mathbf{r}_i) \Delta V = \frac{\omega I_0 \Delta V}{n_m c} \text{Im}(\epsilon_k) c_k,
\label{eq:heat_power_np}
\end{equation}
where $\mathcal{M}_k$ denotes the set of cells belonging to nanoparticle $k$, and the integrated field coefficient is defined as:

\begin{equation}
c_k = \sum_{i \in \mathcal{M}_k} \left|\frac{\mathbf{E}(\mathbf{r}_i)}{\mathbf{E}_0}\right|^2.
\label{eq:field_coefficient}
\end{equation}

\subsection{Temperature calculation via LMI}

The steady-state temperature distribution was computed by solving the heat diffusion equation:

\begin{equation}
\kappa_m \nabla^2 T(\mathbf{r}) = -q(\mathbf{r}),
\label{eq:heat_diffusion}
\end{equation}
where $\kappa_m$ is the thermal conductivity of the medium. For metallic nanoparticles in aqueous environments ($\kappa_{\text{Au}},\kappa_{\text{Ag}} \gg \kappa_{\text{H}_2\text{O}}$), the temperature within each nanoparticle is approximately uniform, following the Uniform Temperature Approximation \cite{Baffou2010}.

To determine the temperature of each of the $P$ nanoparticles in a system, we employed the LMI  formalism. While the original work derived the formalism for a single structure and provided the specific coupled equations for a dimer (two disjointed structures), we generalize it here for an arbitrary number $P$ of nanoparticles.

The uniform temperatures of the nanoparticles, represented by the vector $\mathbf{T} = [T_1, T_2,$ $\ldots, T_P]$, are obtained by solving the following coupled linear system derived from energy conservation and the LMI formalism:

\begin{equation}
\begin{pmatrix}
c_1 \\
c_2 \\
\vdots \\
c_M
\end{pmatrix}
= \frac{4\pi \kappa_m n_m c}{\omega I_0 \text{Im}(\epsilon) \Delta V}
\begin{pmatrix}
r_{L1} & r_{12} & \cdots & r_{1P} \\
r_{21} & r_{L2} & \cdots & r_{2P} \\
\vdots & \vdots & \ddots & \vdots \\
r_{P1} & r_{P2} & \cdots & r_{LP}
\end{pmatrix}
\begin{pmatrix}
T_1 \\
T_2 \\
\vdots \\
T_P
\end{pmatrix},
\label{eq:coupled_system}
\end{equation}
where $\kappa_m = 0.598$ W/m$\cdot$K is the thermal conductivity of water, $r_{Lk}$ is the Laplace radius of nanoparticle $k$, and $r_{kl}$ are the coupled radius terms between nanoparticles $k$ and $l$ ($l\in P$ ), both radius defined as:

\begin{align}
r_{Lk} &= \sum_{i,j \in \mathcal{M}_k} (\mathbf{A}^{-1})_{ij}, \\
r_{kl} &= \sum_{i \in \mathcal{M}_k} \sum_{j \in \mathcal{M}_l} (\mathbf{A}^{-1})_{ij}.
\end{align}

The Laplace matrix $\mathbf{A}$ was constructed with elements:

\begin{equation}
A_{ij} = 
\begin{cases}
\dfrac{1}{|\mathbf{r}_i - \mathbf{r}_j|}, & i \neq j \\
\dfrac{1}{R_{\text{eff}}}, & i = j
\end{cases}
\label{eq:laplace_matrix}
\end{equation}
where $R_{\text{eff}} = (3\Delta V/4\pi)^{1/3}$ is the effective radius of a cubic cell. The matrix in Eq. \ref{eq:coupled_system} is symmetric with $r_{kl} = r_{lk}$ since $\mathbf{A}$ is symmetric. For the case $P=2$, this system reduces to the pair of equations (38) and (39) presented in Ref. \citenum{Baffou2010_GreenF}. Our formulation thus provides a direct and general framework to compute the temperature distribution in clusters composed of any number of optically and thermally interacting plasmonic nanoparticles.

\subsection{Fictive source distribution and temperature reconstruction}

To enforce uniform temperature within each nanoparticle, a fictive electric field enhancement distribution $e_i^2$ was computed that satisfies:

\begin{equation}
e_i^2 = \frac{4\pi \kappa_m n_m c}{\omega I_0 \text{Im}(\epsilon) \Delta V} \sum_{j=1}^{M} (\mathbf{A}^{-1})_{ij} T_k,
\label{eq:fictive_field}
\end{equation}
where $T_k$ is the uniform temperature of the nanoparticle that cell $j$ belongs to, as determined from the solution of the coupled system in Eq.~\ref{eq:coupled_system}.

The complete temperature distribution over the $i\in N$ cells was then reconstructed throughout the computational domain using the superposition principle:

\begin{equation}
T(\mathbf{r}_i) = \sum_{j=1}^{M} G(\mathbf{r}_i, \mathbf{r}_j) \, q_{\text{fictive}}(\mathbf{r}_j) \, \Delta V,
\label{eq:temperature_reconstruction}
\end{equation}
where the thermal Green's function is defined as:
\begin{equation}
G(\mathbf{r}_i, \mathbf{r}_j) = 
\begin{cases}
\dfrac{1}{4\pi \kappa_m |\mathbf{r}_i - \mathbf{r}_j|}, & \mathbf{r}_i \neq \mathbf{r}_j \\
\dfrac{1}{4\pi \kappa_m r_{Lj}}, & \mathbf{r}_i = \mathbf{r}_j
\end{cases}
\end{equation}
with $r_{Lj}$ being the Laplace radius that accounts for the self-interaction term, and the fictive heat source density is:

\begin{equation}
q_{\text{fictive}}(\mathbf{r}_j) = \frac{\omega I_0}{n_m c} \text{Im}(\epsilon) \, e_j^2.
\end{equation}

The validity of this methodology was confirmed by reproducing key benchmark cases from the literature, as shown in Figure \ref{SIfig:Validation} of the Supplementary Information, along with key theoretical results such as the characteristic $1/r$ temperature decay outside nanoparticles \cite{Baffou2010}. While more fundamental approaches exist that model electron–phonon interactions and interface effects, including those characterized by the Kapitza length \cite{Cunha2020, Baffou2011}, our approach provides reliable thermal predictions for complex plasmonic nanostructures.

\subsection{Treatment of silica-encapsulated nanoparticle clusters in water}

The cluster system consisted of seven gold nanoparticles encapsulated within a porous silica shell, with the entire system immersed in water, system experimentally achieved in Ref. \citenum{Snch2018}.

Considering that water is the dominant continuous phase both inside and outside the silica shell, and permeates through the porous silica, the effective optical and thermal properties were calculated with water as the matrix phase.

The effective dielectric constant for the silica-water composite was calculated using Bruggeman's effective medium approximation \cite{Bruggeman}, which provides a symmetric treatment of both components:

\begin{equation}
(1-f)\frac{\epsilon_{\text{H}_2\text{O}} - \epsilon_{\text{eff}}}{\epsilon_{\text{H}_2\text{O}} + 2\epsilon_{\text{eff}}} + f\frac{\epsilon_{\text{SiO}_2} - \epsilon_{\text{eff}}}{\epsilon_{\text{SiO}_2} + 2\epsilon_{\text{eff}}} = 0,
\label{eq:bruggeman_effective_opt}
\end{equation}
where $\epsilon_{\text{H}2\text{O}} = 1.777$, $\epsilon_{\text{SiO}_2} = 2.103$ (the value used in Ref. \citenum{Snch2018} corresponding to fused silica), and $f$ is the volume fraction of silica in the cluster. For a shell with outer radius $R_{\text{outer}} = 158.5$ nm and thickness $t = 20$ nm, the silica volume fraction is:

\begin{equation}
f = \frac{R_{\text{outer}}^3 - (R_{\text{outer}} - t)^3}{R_{\text{outer}}^3} \approx 0.33.
\end{equation}

Solving equation~\ref{eq:bruggeman_effective_opt} numerically yields $\epsilon_{\text{eff}} \approx 1.881$ for the effective medium surrounding the gold nanoparticles.

The effective thermal conductivity was calculated using Bruggeman's approximation to maintain consistency with the optical modeling:

\begin{equation}
(1-f)\frac{\kappa_{\text{H}_2\text{O}} - \kappa_{\text{eff}}}{\kappa_{\text{H}_2\text{O}} + 2\kappa_{\text{eff}}} + f\frac{\kappa_{\text{SiO}_2} - \kappa_{\text{eff}}}{\kappa_{\text{SiO}_2} + 2\kappa_{\text{eff}}} = 0,
\label{eq:bruggeman_effective_Tcond}
\end{equation}
where $\kappa_{\text{SiO}_2} = 1.4$ W/mK\cite{WYPYCH201613}. Solving equation \ref{eq:bruggeman_effective_Tcond} numerically yields $\kappa_{\text{eff}} \approx 0.81$, which is still much smaller than the thermal conductivities of the metallic nannoparticles.

\section{Results}

The optical and thermal properties of metallic nanoparticles are governed by the localized surface plasmon resonance (LSPR), a collective oscillation of the conduction electrons excited by an incident light at specific wavelength \cite{Kelly2003,Willets2007}. Upon excitation, a portion of the absorbed energy is re-radiated via scattering, while the remainder is dissipated non-radiatively, primarily as heat; this efficient light-to-heat conversion is the very basis of thermoplasmonics. Furthermore, since the LSPR is highly sensitive to the nanoparticle's size, shape, material, and local environment, these parameters provide a direct means to control its thermal behavior. 

\begin{figure}
  \includegraphics[width=16 cm]{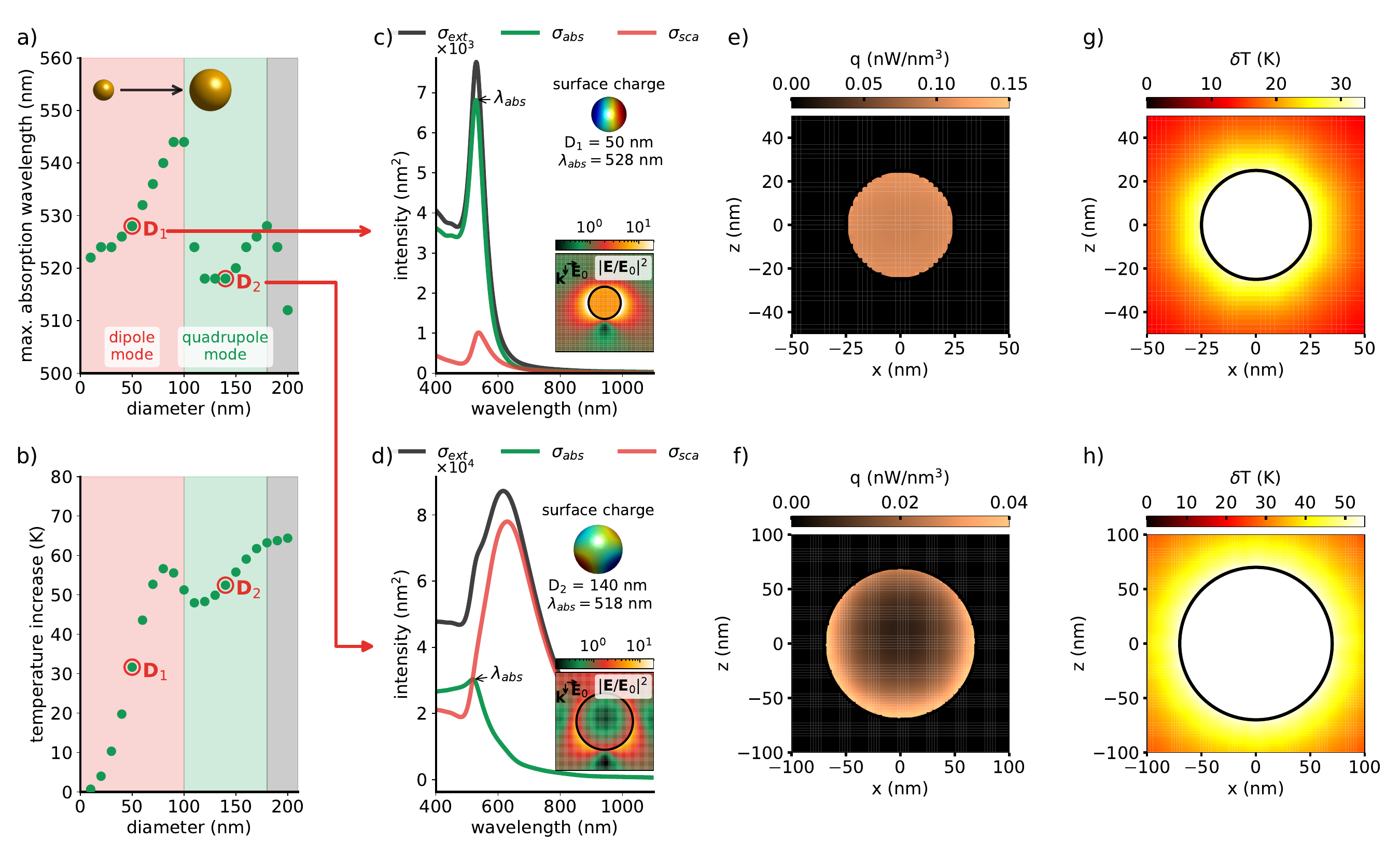}
  \caption{Thermoplasmonic response of an individual Au nanoparticle: diameter dependence. (a) Maximum absorption wavelength ($\lambda_{abs}$) as a function of nanoparticle diameter. The shaded regions indicate the dominant plasmonic modes: dipole (red), quadrupole (green), and octupole (black). (b) Temperature increase ($\delta$T) of the nanoparticle at $\lambda_{abs}$ as a function of its diameter (excitation irradiance: $10^9$ W/m$^2$). (c, d) Extinction, absorption, and scattering cross-section spectra for nanoparticles with diameters of 50 nm (c) and 140 nm (d). Insets show the corresponding surface charge distribution and electric field maps in the plane of vibration (defined by the wave vector and E-field polarization) at $\lambda_{abs}$. (e, f) Heat power density (q) maps in the plane of vibration for the 50 nm (e) and 140 nm (f) nanoparticles. (g, h)  Steady-state temperature profiles for the 50 nm (g) and 140 nm (h) nanoparticles.}
  \label{Fig:AuNP_single_res}
\end{figure}

The thermoplasmonic response of a single gold nanosphere (Figure \ref{Fig:AuNP_single_res}) exhibits a strong size dependence. With increasing diameter, the LSPR redshifts (Figure \ref{Fig:AuNP_single_res}a) and the absorption cross-section intensifies (Figure \ref{SIfig:CS_single_AuNP}) as the dominant plasmonic mode transitions from dipole to higher-order modes. Consequently, the resulting temperature rise is a function of both excitation wavelength and particle size. Figure \ref{Fig:AuNP_single_res}b shows that the temperature increase reaches up to 60 K for diameters below 100 nm when excited at their respective absorption maxima ($\lambda_{abs}$).
Contrary to the intuition that a larger metal volume invariably leads to more heating, the temperature increase does not scale monotonically. For diameters beyond approximately 80 nm, the temperature profile exhibits a non-trivial pattern of decreases and subsequent increases. This complex behavior is a direct consequence of the shifting dominance between different plasmonic modes.

The nature of the dominant plasmonic mode also dictates the spatial distribution of the enhanced electric field in the nanoparticle's near-field (Figure \ref{SIFig:AuNP_Efield_single_res}), a critical factor that will influence the plasmonic coupling with neighboring nanostructures. Then, the temperature behavior will be also affected by the distance, positions, and composition. Furthermore, the efficiency of heat generation is intrinsically linked to the balance between absorption and scattering. For a small nanoparticle (50 nm, Figure \ref{Fig:AuNP_single_res}c) dominated by the dipole mode, absorption prevails over scattering at the LSPR, making it a highly efficient nano-heater. In contrast, for a larger nanoparticle (140 nm, Figure \ref{Fig:AuNP_single_res}d), scattering becomes the dominant decay pathway at the LSPR, significantly reducing its heating efficiency. It is crucial to note that the wavelength of maximum extinction, commonly measured in UV-Vis-NIR spectroscopy, does not always coincide with the wavelength of maximum absorption ($\lambda_{abs}$), where the peak of temperature occurs. This underscores the necessity of numerical calculations or advanced experimental measurements to determine the optimal heating conditions in complex systems.

The heat power density profile within the nanoparticle reflects the underlying electric field distribution (Figures \ref{Fig:AuNP_single_res}e and \ref{Fig:AuNP_single_res}f; insets of Figures \ref{Fig:AuNP_single_res}c and \ref{Fig:AuNP_single_res}d). While the 50 nm sphere exhibits a relatively homogeneous heat source, the 140 nm sphere shows a highly non-uniform profile concentrated at its surface. Nevertheless, under the Uniform-Temperature Approximation, the nanoparticle itself is considered isothermal, with the resulting steady-state temperature field decaying radially outward from its surface (Figures \ref{Fig:AuNP_single_res}g and \ref{Fig:AuNP_single_res}h). While we focus on gold here, similar size-dependent trends are observed for silver nanoparticles (Figure \ref{SIFig:AgNP_single_res}), albeit with LSPRs shifted to lower wavelengths due to their different dielectric function. 

\begin{figure}
  \includegraphics[width=12 cm]{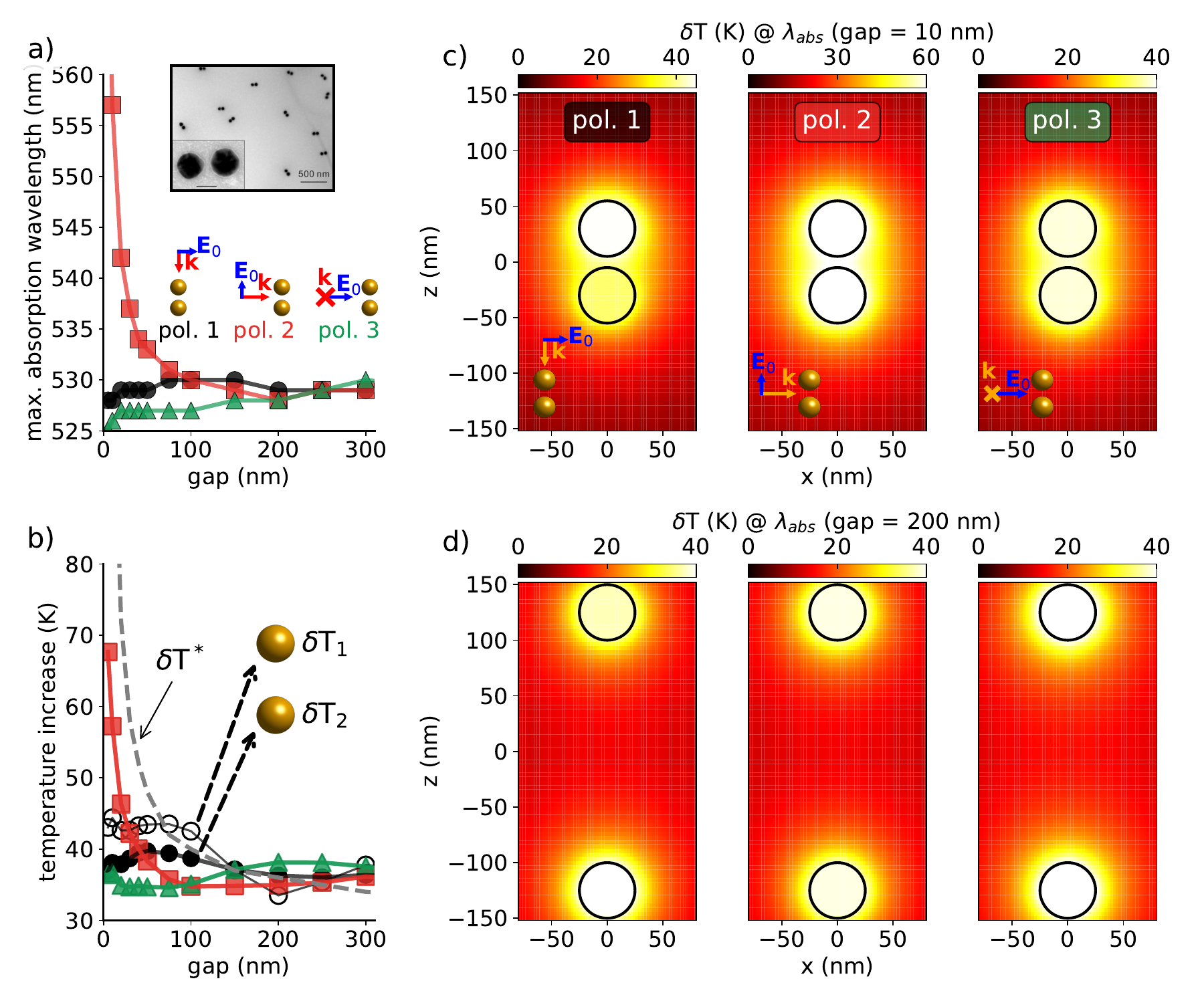}
  \caption{Thermoplasmonic response of Au nanoparticle dimers: gap distance and polarization dependence. (a) Maximum absorption wavelength as a function of interparticle gap for three E-field polarizations: transverse to the dimer axis (pol. 1), longitudinal along the dimer axis (pol. 2), and at 90$^0$ orientation to the axis (pol. 3). The inset shows an illustrative TEM micrograph of dimers enabled by the DNA origami technique (reproduced from Ref. \citenum{Xin2019}; Copyright 2019 American Chemical Society). (b) Temperature increase of each nanoparticle ($\delta$T$_1$, solid marks, and $\delta$T$_2$, hollow marks) at $\lambda_{abs}$ as a function of the gap for each polarization (as indicated in (a)). $\delta T^*=Q(1+a/d)/4\pi\kappa_ma$ (dashed line) is an approximation of collective temperature of each nanoparticle with radius $a$ and gap $d$ delivering equal heat power $Q$ (calculated for a single nanoparticle, see Ref. \citenum{Baffou_2017}). Excitation irradiance: $10^9$ W/m$^2$. (c, d) Steady-state temperature profiles for dimers with gaps of 10 nm (c) and 200 nm (d).}
  \label{Fig:AuNP_dimer_res}
\end{figure}

Next, we investigate dimers comprising 50 nm gold nanospheres. Such structures can be precisely assembled using bottom-up techniques like DNA origami scaffolds, which enables controlled nanoscale gaps (see Figure \ref{Fig:AuNP_dimer_res}a, inset) \cite{Kuzyk2018, Xin2019}. While the high symmetry of a single sphere makes its response polarization-independent,  the dimer geometry breaks spherical symmetry, resulting in an axisymmetric structure ($D_{\infty h}$ point group). Consequently, its optical and thermal response strongly depends on the direction of the incident electric field relative to the dimer axis. 
We systematically examine how the thermoplasmonic response is modulated by the interparticle gap distance for the three key orthogonal polarization directions: transverse to the dimer axis, longitudinal along
the dimer axis, and at 90$^0$ orientation to the axis, as illustrated in Figure \ref{Fig:AuNP_dimer_res}.

The optical coupling in a dimer system is modulated by its interparticle gap and the incident polarization, which together control the coupling strength and the hybridized plasmon mode. As illustrated in Figure \ref{Fig:AuNP_dimer_res}a, the maximum absorption wavelength ($\lambda_{abs}$) for large gaps approaches the single nanoparticle resonance, almost polarization-independent, confirming the absence of significant coupling. As the gap decreases, the resonance behavior diverges sharply based on polarization. For the transverse (Pol. 1) and 90$^0$ (Pol. 3) polarizations, a slight blue-shift of $\lambda_{abs}$ is observed. In contrast, the longitudinal polarization (Pol. 2) induces a strong red-shift, which becomes pronounced for gaps below 50 nm. This is a classic signature of plasmon hybridization, where the longitudinal excitation forms a bonding dipole plasmon mode \cite{Nordlander2004}. The complete optical cross-section spectra (Figure \ref{SIfig:CS_dimer_AuNP}) reveal that for Pol. 1 and Pol. 3, absorption remains the dominant process over scattering even at small gaps. For Pol. 2, however, the scattering cross-section increases significantly at small gaps.

This polarization-dependent coupling directly dictates the thermal response. Figure \ref{Fig:AuNP_dimer_res}b plots the temperature increase of each nanoparticle ($\delta$T$_1$, $\delta$T$_2$) at their respective $\lambda_{abs}$ as a function of the gap. For large gaps, the temperature for all polarizations converges to the single-particle value of $\sim$33 K. As the gap decreases, the temperature generally rises due to the proximity of the two heat sources. However, for the longitudinally polarized case (Pol. 2) at sub-50 nm gaps, the temperature increases dramatically. This trend is compared to an approximated model, $\delta$T$^*$ (dashed line in \ref{Fig:AuNP_dimer_res}b) \cite{Baffou2010}, which calculates the temperature by summing the contributions of individual, non-coupled nanoparticles. The observed deviation, where $\delta$T$^*$ overestimates the actual temperature, arises because the model fails to account for the increased radiative scattering in the strongly coupled dimer.
 
A key finding is the emergence of asymmetric heating under transverse excitation (Pol. 1), whereas the other two polarizations do not produce significant temperature differences between nanoparticles. Under Pol. 1, the top nanoparticle ($\delta$T$_1$) consistently reaches a higher temperature than the bottom one ($\delta$T$_2$), with a difference of approximately $\sim5$ K at a 10 nm gap (Figure \ref{Fig:AuNP_dimer_res}c). This asymmetry persists, though inverted, at a 200 nm gap ($\sim1.5$ K, Figure \ref{Fig:AuNP_dimer_res}d), indicating a complex interplay between near-field shadowing and far-field interference that is dominant under this transverse polarization \cite{Baffou2010}. Ultimately, this establishes a localized thermal gradient that could, in principle, be harnessed in electrically-connected systems to induce a Seebeck-like thermoelectric response, pointing towards applications in plasmon-enabled energy conversion.

The underlying mechanisms for these thermal profiles are revealed by the electric field and charge density distributions (Figures \ref{SIFig:AuNP_Efield_dimer_gsmall_res} and \ref{SIFig:AuNP_sigma_dimer_gsmall_res} for a small gap, \ref{SIFig:AuNP_Efield_dimer_gbig_res} and \ref{SIFig:AuNP_sigma_dimer_gbig_res} for large gap). For a small gap under Pol. 1, the incident wave (k-vector from top) induces a strong dipolar mode in the top nanoparticle, which effectively ``shadows" the bottom particle. This leads to the formation of a higher-order mode (quadrupole) in the bottom nanoparticle, as confirmed by the charge density. The resultant higher absorption in the top nanoparticle explains the observed temperature asymmetry. For Pol. 2 at a small gap, the characteristic bonding dipole plasmon mode is established, concentrating immense electric field intensity in the gap and leading to strongly enhanced local heating. Regarding Pol. 3 at a small gap, the dipoles oscillate in-phase in an unbounded configuration. The resulting charge screening between the nanoparticles suppresses near-field coupling, leading to absorbed fields and temperature increases comparable to the large-gap scenario.
When the gap is large, the optical coupling vanishes, and both particles exhibit simple, uncoupled dipolar excitations with temperatures similar to single nanoparticles. 

In summary, the dimer system demonstrates that optical coupling introduces critical degrees of freedom, namely, polarization and gap distance. These allow for the tuning of the resonance wavelength and provide control over the magnitude and spatial distribution of heat generation.

\begin{figure}
  \includegraphics[width=16 cm]{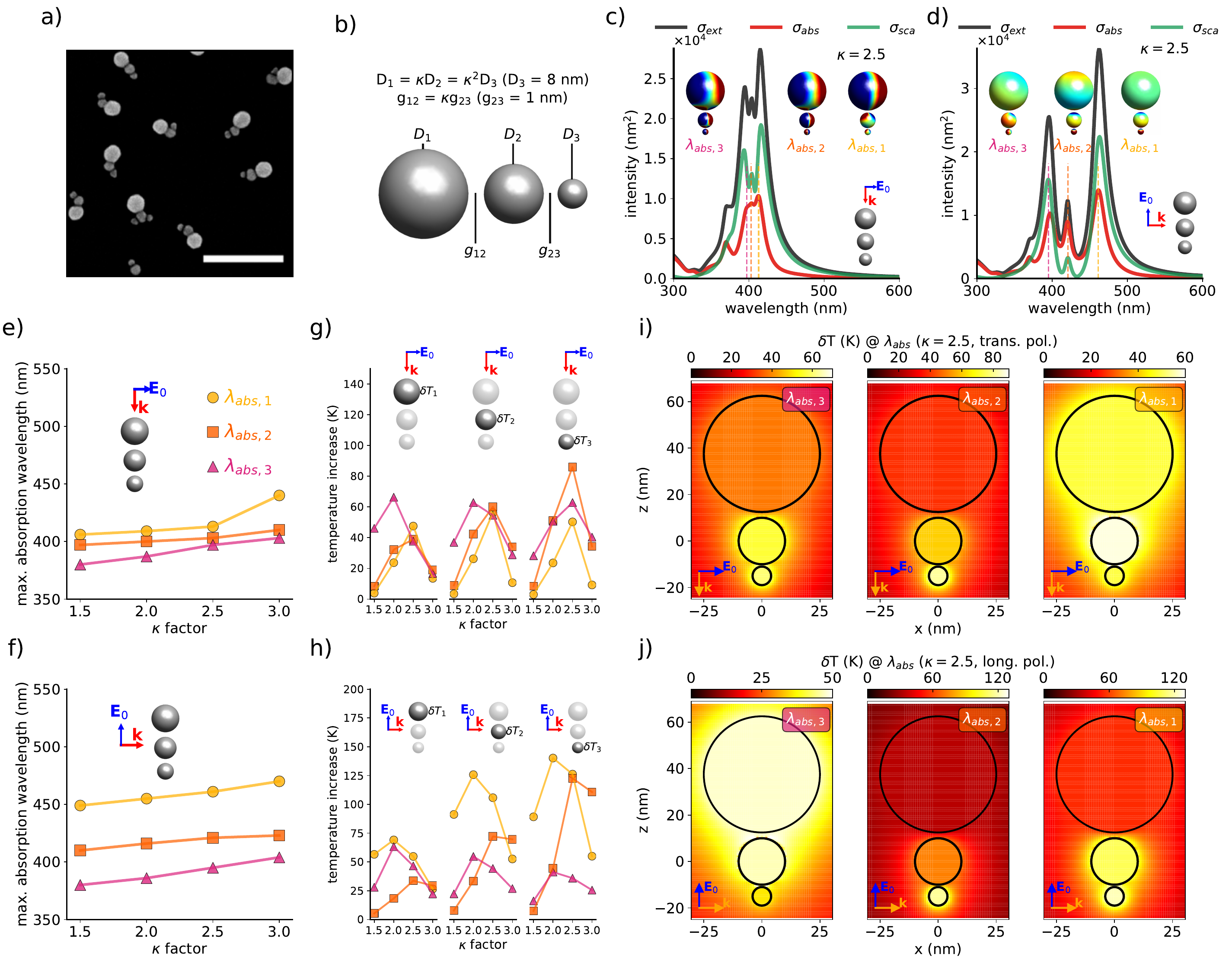}
  \caption{Thermoplasmonic response of Ag nanoparticle nanolens: size ratio and polarization dependence. (a) Experimental TEM micrograph of trimer nanolenses (reproduced with permission from Ref.  \citenum{Lloyd2017}; Copyright 2017 American Chemical Society). (b) Scheme of the self-similar nanolens, where D$_1$, D$_2$, and D$_3$ are the nanoparticle diameters, g$_{12}$ and g$_{23}$ are the gap between D$_1$-D$_2$ and D$_2$-D$_3$, respectively, and $\kappa$ is the self-similar factor. All the results are for fixed D$_3 = 8$ nm and g$_{23} = 1$ nm. (c, d) Extinction, absorption, and scattering cross-section spectra of a nanolens with $\kappa=2.5$ under transverse (c) and longitudinal (d) excitations. Insets: Surface charge distributions at the three principal absorption resonances $\lambda_{abs,i}$ ($i=1,2,3$, as indicated by the vertical dashed lines). (e, f) Evolution of the resonance absorption wavelengths with the self-similar factor $\kappa$ for transverse (e) and longitudinal (f) polarizations. (g, h) Temperature increase of individual nanoparticles ($\delta$T$_1$, $\delta$T$_2$, and $\delta$T$_3$) at $\lambda_{abs,i}$ as a function of $\kappa$, for both transverse (g) and longitudinal (h) polarizations. Excitation irradiance: $10^9$ W/m$^2$. (i, j) Steady-state temperature profiles for nanolens with $\kappa=2.5$ at $\lambda_{abs,i}$ under transverse (i) and longitudinal (j) polarizations.}
  \label{Fig:AgNP_nanolens_res}
\end{figure}

The principles of asymmetric and enhanced heating established in the dimer structure can be strategically engineered in more complex architectures. We now explore a plasmonic nanolens, a structure recognized for its ability to focus electromagnetic energy into deep sub-wavelength volumes, far beyond the diffraction limit. The concept, originally proposed as a self-similar chain of metal nanospheres by Li \textit{et al}.\cite{Li2003}, relies on cascaded near-field interactions to concentrate light. Such structures have been successfully realized experimentally, for example, through electrostatic self-assembly methods \cite{Lloyd2017}, as shown in Figure \ref{Fig:AgNP_nanolens_res}a.

This system allows us to investigate whether such field focusing translates into a `nano-heat-lens' effect, concentrating thermal energy into a deeply sub-wavelength volume. To explore this, we investigate a silver nanolens trimer. Silver is chosen over gold for this study due to its lower ohmic losses, which results in narrower, more distinguishable resonance peaks essential for resolving the complex coupling in this system. We model a linear trimer with geometrically scaled dimensions (see Figure \ref{Fig:AgNP_nanolens_res}b): the diameters follow $D_1=\kappa D_2=\kappa^2D_3$, and the gaps follow $g_{12}=\kappa g_{23}$, where $D_3=8$ nm and $g_{23}=1$ nm are fixed. The self-similar factor $\kappa$ is varied (1.5, 2, 2.5, 3) to systematically tune the optical response.

Given the axial symmetry of the linear trimer, we analyze the two principal incident polarizations: transverse and longitudinal to the chain axis. The extinction spectra for a nanolens with $\kappa=2.5$ (Figures \ref{Fig:AgNP_nanolens_res}c and \ref{Fig:AgNP_nanolens_res}d) reveal three distinct absorption peaks ($\lambda_{abs,1}$, $\lambda_{abs,2}$, $\lambda_{abs,3}$) for each polarization, corresponding to hybridized plasmonic modes of the structure (see full cross-section spectra in Figure \ref{SIfig:CS_nanolens_AgNP}). As $\kappa$ increases, the nanoparticles grow in size, causing a systematic red-shift of all resonance wavelengths (Figures \ref{Fig:AgNP_nanolens_res}e and \ref{Fig:AgNP_nanolens_res}f).

The thermoplasmonic response, characterized by the temperature increase of each nanoparticle at its respective  $\lambda_{abs,i}$, is complex and highly tunable (Figures \ref{Fig:AgNP_nanolens_res}g and \ref{Fig:AgNP_nanolens_res}h). A key trend is that temperatures generally increase with $\kappa$ up to an optimum value of $\kappa=2.5$, beyond which they decrease due to a reduction in absorption efficiency, mirroring the size-effects observed in single nanoparticles. While the absolute maximum temperature ($\sim140$ K) occurs for $\kappa=2$ at $\lambda_{abs,1}$, longitudinal polarization, we identify $\kappa=2.5$ as the optimal design parameter based on the most pronounced thermal gradient between nanoparticles, with temperatures of 122 K, 72 K, and 33 K for the small, medium, and large nanoparticles, respectively, at $\lambda_{abs,2}$. This steep thermal gradient, where the smallest nanoparticle reaches temperatures more than an order of magnitude higher than an isolated 8 nm Ag sphere ($\sim10$ K, see Figure \ref{SIFig:AgNP_single_res}), more effectively demonstrates the nano-heat-lensing effect for thermal energy concentration.

The underlying coupling mechanisms are elucidated by the electric field and charge distributions for $\kappa=2.5$ (Figures \ref{SIFig:AgNP_Efield_nanolens_res}) and \ref{SIFig:AgNP_sigma_nanolens_res}. Under transverse polarization, excitation at $\lambda_{abs,2}$ induces a cascade of dipole-mediated bonding modes that most effectively focus the electric field into the $g_{23}$ gap, yielding the highest temperature ($\sim86$ K) in the smallest nanoparticle (Figure \ref{Fig:AgNP_nanolens_res}i). For longitudinal polarization, a similar but stronger effect is observed due to the direct field enhancement along the axis. Excitation at $\lambda_{abs,2}$ establishes a fully coupled bonding dipole mode across all three nanoparticles, creating an intense hot spot and achieving a peak temperature of 122 K in $D_3$ (Figure \ref{Fig:AgNP_nanolens_res}j). It is worth noting that at this resonance, the absorption cross-section dominates over scattering, resulting in a heating efficiency that is significantly higher than in the other modes, where scattering plays a more dominant role (Figure \ref{Fig:AgNP_nanolens_res}d). 
At other resonances, the mode hybridization is less effective at focusing energy exclusively into the smallest particle (as in the case of Figure \ref{Fig:AgNP_nanolens_res}d for $\lambda_{abs,1}$ or $\lambda_{abs,3}$), leading to reduced nanolens efficiency or a complete suppression of the effect.

In conclusion, the silver nanolens demonstrates that cascaded near-field coupling can be engineered to concentrate not only electric fields but also thermal energy into a nanoscale volume, creating an ultra-localized heat source. The performance is highly sensitive to the geometric factor $\kappa$ and the excitation wavelength, with an optimal configuration identified here. Navigating this multi-parameter design space is challenging, but emerging algorithm-driven optimization tools promise to accelerate the finding of optimal geometries for specific thermoplasmonic applications \cite{JLMP2025}.

\begin{figure}
  \includegraphics[width=16 cm]{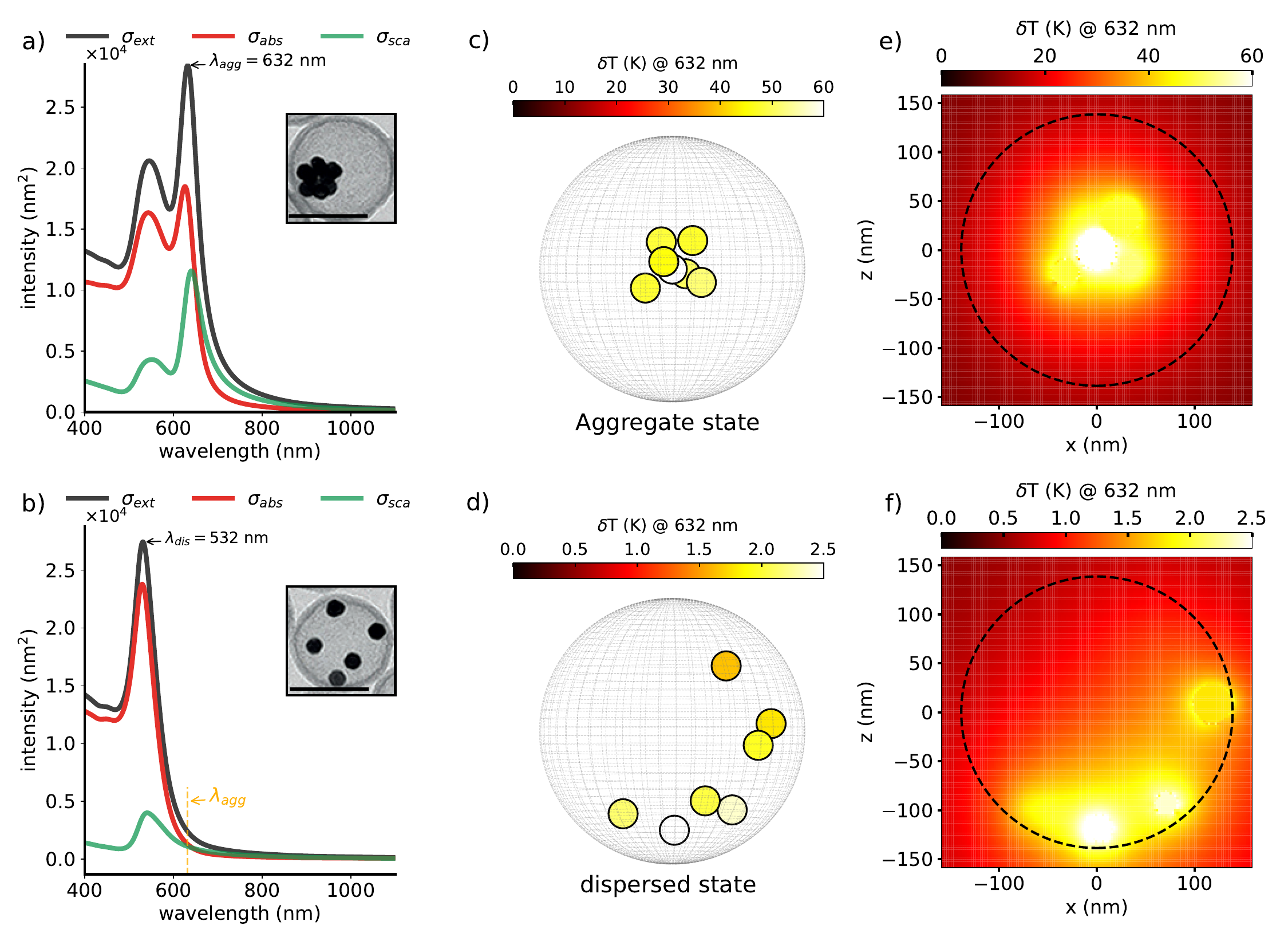}
  \caption{Thermoplasmonic response of gold nanoparticle clusters confined within a silica shell: aggregated and dispersed states. (a, b) Extinction, absorption, and scattering cross-section spectra for aggregated (a) and dispersed (b) nanoparticle arrangements. Each cluster consists of seven Au nanoparticles on average within a silica shell. The spectra result from averaging over 15 distinct random nanoparticle positions for each state. Insets show experimental TEM micrographs of Au nanoparticle clusters in the two states (adapted from Ref. \citenum{Snch2018}; reproduced with permission). (c, d) Increase of temperature of individual nanoparticles for representative aggregated (c) and dispersed (d) configurations, excited at $\lambda_{agg}=632$ nm. Excitation irradiance: $10^9$ W/m$^2$. (e, f) Steady-state temperature profiles (xz-plane at y = 0 nm) for the aggregated (e) and dispersed (f) states at $\lambda_{agg}=632$ nm. Dashed black line indicates the shell inner surface}
  \label{Fig:AgNP_cluster_res}
\end{figure}

We now extend our study to a dynamic cluster of gold nanoparticles confined within a capsule, where the optical and thermal properties can be actively switched by external stimulus. The concept, illustrated in Figure \ref{Fig:AgNP_cluster_res}, a switchable cluster under confinement.

The model is based on our past experimental systems of polystyrene-functionalized gold nanoparticles encapsulated in a mesoporous silica shell, permeable to solvent molecules.\cite{Snch2018} The reversible clustering is driven by solvent composition, where low-polarity solvent maintains nanoparticles dispersed while high-polarity solvent leads to aggregation due to hydrophobic forces. Each cluster consists of seven Au nanoparticles (40 nm) within a silica shell of 20 nm thick and inner diameter of 277 nm. Experimental TEM images of the aggregated and dispersed states are shown in the insets of Figures \ref{Fig:AgNP_cluster_res}a and \ref{Fig:AgNP_cluster_res}b, respectively.
In our simulations, the aggregated state features a minimum interparticle gap of 2 nm. The optical cross-section spectra (Figures \ref{Fig:AgNP_cluster_res}a and \ref{Fig:AgNP_cluster_res}b) reveal a broad, red-shifted extinction band at $\lambda_{agg}=632$ nm for the aggregate state due to strong near-field coupling, while the dispersed state shows a blue-shifted band at 
$\lambda_{dis}=532$ nm, characteristic of uncoupled dipoles.

We calculated the thermal response under fixed excitation at $\lambda_{agg}=632$ nm. Figure \ref{Fig:AgNP_cluster_res}c shows the temperature distribution for a representative aggregated configuration, where the strong optical coupling leads to a significant temperature increase of approximately 60 K. In contrast, the same excitation of the dispersed state (Figure \ref{Fig:AgNP_cluster_res}d) yields a temperature rise of $\sim2.5$ K per nanoparticle, as the excitation of the system is far from resonance. 
This remarkable $\sim$60 K temperature difference between the two states reveals the potential for a self-regulating system, where the optical and thermal states are intrinsically coupled. The steady-state temperature profiles (Figures \ref{Fig:AgNP_cluster_res}e and \ref{Fig:AgNP_cluster_res}f) further highlight this contrast, showing intense, localized heating in the aggregated core and a mild, distributed temperature field in the dispersed case.

Although the localized thermal profiles presented here are calculated for an idealized steady-state scenario (nanoparticles in an infinite, homogeneous water medium), prolonged continuous illumination in a real, finite system would lead to cumulative heat diffusion, eventually homogenizing the temperature field in the microenvironment, yielding to the degradation of thermal gradients. To preserve these focused thermal profiles (such as the intense heat in the smallest nanoparticle of the nanolens) the excitation must be applied in pulses shorter than the characteristic heat diffusion time. This timescale is inherently transient; for a nanoscale heat source in water, the thermal evolution occurs on the order of nanoseconds. Indeed, following the framework of Baffou \textit{et al}. \cite{Baffou_2017}, a plasmonic system can reach $\sim84\%$ of its final temperature increase in times as short as 10 ns, with a comparably rapid decay (e.g., 20–50 ns) once illumination ceases.
Therefore, our findings indicate that the experimental observations of the proposed here nano-heat spots requires a pulsed excitation to balance incident power and pulse width. A higher incident power enables the target temperature to be reached faster, thereby permitting the use of shorter pulses to effectively confine the heat. Optimizing this pulse-power interplay is essential to achieve intense, localized heating while mitigating global heat accumulation in the microenvironment, constituting a key design principle for thermoplasmonics under optically-coupled regime.

\section{Conclusions}

In summary, inspired by experimentally achievable plasmonic nanostructures we provided a numerical design principle for achieving precise spatiotemporal control over nanoscale heat generation under optical coupling. We have demonstrated that by engineering near-field interactions (through parameters like gap distance, polarization, and hierarchical geometry) it is possible to dictate not just the magnitude, but also the localization and symmetry of photothermal heating. This control ranges from creating asymmetric thermal profiles in dimers, focused heat spots in nanolenses, and demonstrating large temperature difference ($\sim60$) under reversible states of nanoparticle clusters.

The implications of this work are significant for the future of nanoscale thermal management and actuation. The principles of controlled heating elucidated here, through symmetry breaking, mode hybridization, and structural reconfiguration, can be directly applied to challenges in photothermal catalysis, where precise temperature control at reaction sites could enhance selectivity and yield; in nanofabrication and nano-soldering, where directed heat is required; and in the development of smart, adaptive plasmonic devices. By providing a coherent framework from fundamental single-particle effects to complex, active systems, this study lays the groundwork for designing next-generation thermoplasmonic platforms where heat is not merely a waste, but a finely controlled design element, especially under an optically coupled regime. 

\begin{acknowledgement}
JLM-P acknowledges the financial support received from the IKUR Strategy under the collaboration agreement between the Ikerbasque Foundation and Materials Physics Center on behalf of the Department of Science, Universities and Innovation of the Basque Government. M. G. acknowledges the grant PID2022-141017OB-I00 funded by MCIN/ AEI/10.13039/ 501100011033 and by “ERDF A way of making Europe", grant HRZZ-IP-2022-10-3456 funded by Croatian Science Foundation and grant IT1526-22 funded by Department of Education of the Basque Government.  The authors thank Prof. Guillaume Baffou for his insightful discussions that improved the scientific rigor of this work. The authors acknowledge the technical and human support provided by the DIPC Supercomputing Center, where the simulations were performed.

\end{acknowledgement}

\begin{suppinfo}

Comparison with reported systems (Figure \ref{SIfig:Validation}). Optical cross-sections of single gold nanoparticles (Figure \ref{SIfig:CS_single_AuNP}). Electric field distribution of single gold nanoparticles (Figure \ref{SIFig:AuNP_Efield_single_res}). Thermoplasmonic response for single silver nanoparticles (Figure \ref{SIFig:AgNP_single_res}). Optical cross-sections of single silver nanoparticles (Figure \ref{SIfig:CS_single_AgNP}). Optical cross-sections of gold nanoparticle dimers (Figure \ref{SIfig:CS_dimer_AuNP}). Electric field distribution of gold nanoparticle dimer with small gap (Figure \ref{SIFig:AuNP_Efield_dimer_gsmall_res}). Surface charge density of gold nanoparticle dimer with small gap (Figure \ref{SIFig:AuNP_sigma_dimer_gsmall_res}). Electric field distribution of gold nanoparticle dimer with large gap (Figure \ref{SIFig:AuNP_Efield_dimer_gbig_res}). Surface charge density of gold nanoparticle dimer with large gap (Figure \ref{SIFig:AuNP_sigma_dimer_gbig_res}). Optical cross-sections of silver nanoparticle nanolens (Figure \ref{SIfig:CS_nanolens_AgNP}). Electric field distribution of silver nanoparticle nanolens (Figure \ref{SIFig:AgNP_Efield_nanolens_res}). Surface charge density of silver nanoparticle nanolens (Figure \ref{SIFig:AgNP_sigma_nanolens_res}).

\end{suppinfo}

\bibliography{References}

\end{document}


\clearpage

\begin{figure}
    \centering
    \subfigure[]{\includegraphics[width=0.49\textwidth]{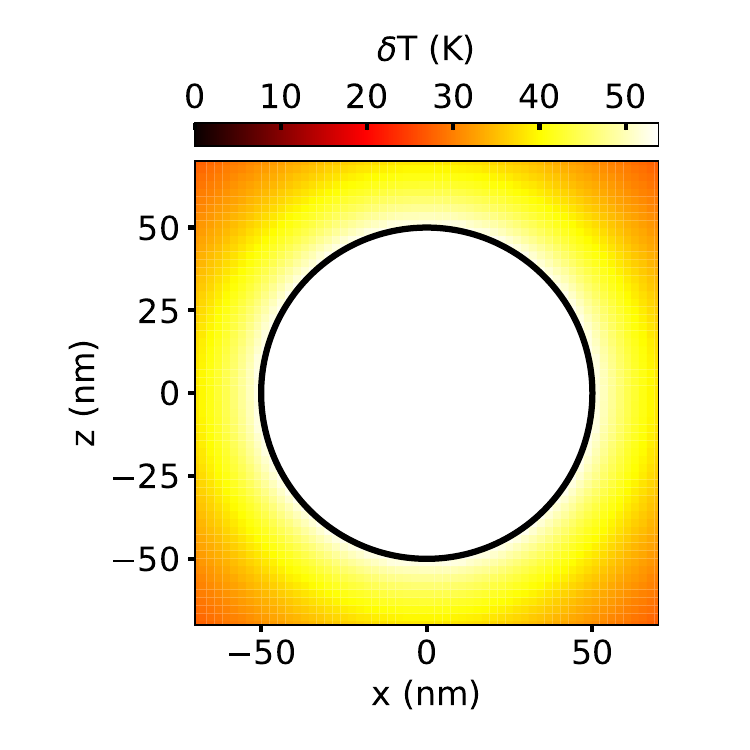}} 
    \subfigure[]{\includegraphics[width=0.49\textwidth]{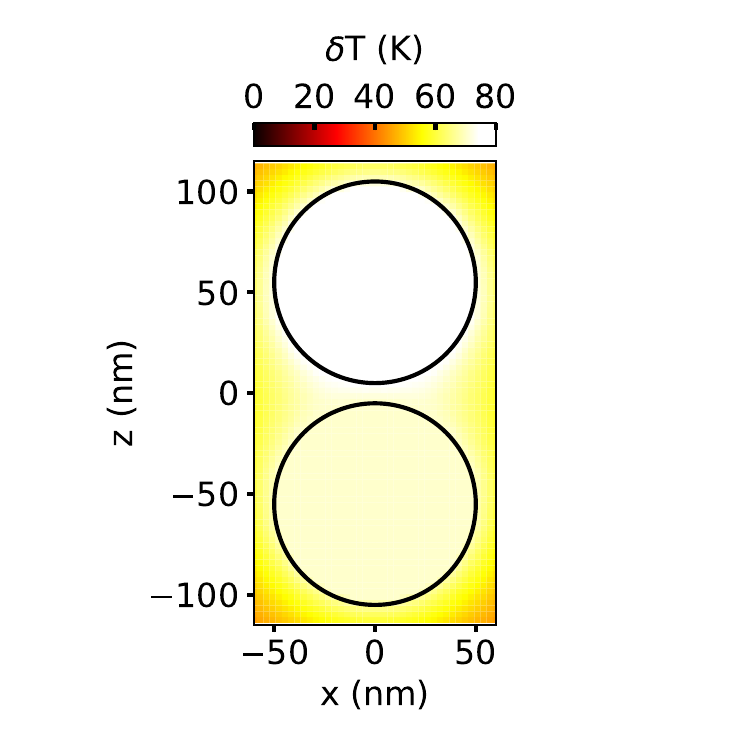}} 
    \caption{Temperature increase of (a) single 100-nm AuNP and (b) a 100-nm-AuNP dimer with a 5-nm gap under 45$^\circ$ incident excitation angle, both in water. The results obtained hear show excellent agreement with those reported in Ref. \citenum{Baffou2010} (their Fig. 1 and Fig. 4, respectively).}
    \label{SIfig:Validation}
\end{figure}

\begin{figure}
  \includegraphics[width=8 cm]{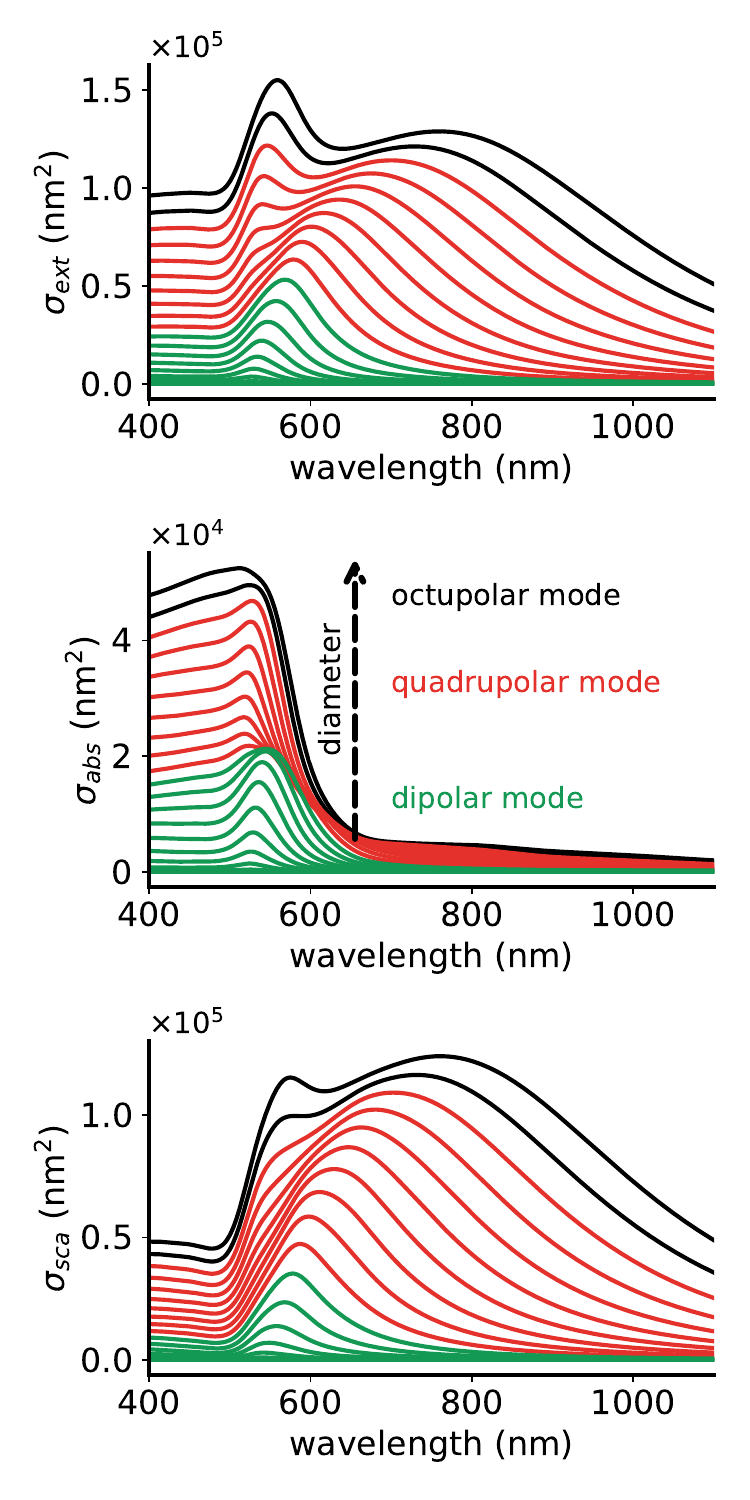}
  \caption{Extinction ($\sigma_{ext}$), absorption ($\sigma_{abs}$), and scattering ($\sigma_{sca}$) cross-section spectra of a single gold nanoparticle with diameter varied from 10 to 200 nm. The color of the spectra (green, red, black) denotes the dominant contributing plasmon resonance: dipole, quadrupole, and octupole, respectively.}
  \label{SIfig:CS_single_AuNP}
\end{figure}

\begin{figure}
  \includegraphics[width=17 cm]{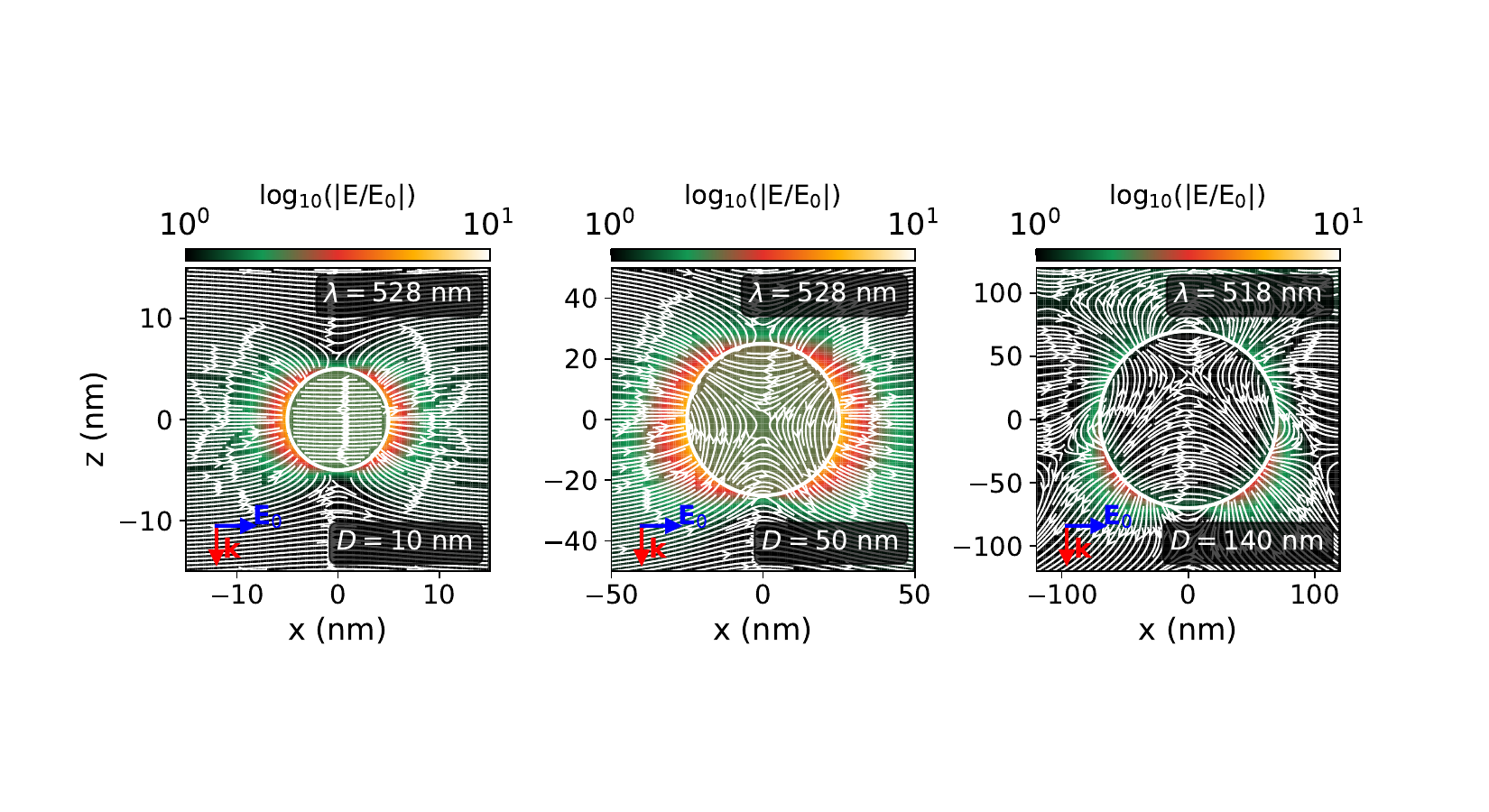}
  \caption{Electric field enhancement ($\text{log}|\textbf{E}/\textbf{E}_0|^2$) and electric field lines in the xz-plane for single gold nanoparticles with diameters of 10 nm, 50 nm, and 140 nm. The white circle represents the nanoparticle boundary. The incident wave vector ($\textbf{k}$) and incident electric field ($\textbf{E}_0$) directions are indicated.}
  \label{SIFig:AuNP_Efield_single_res}
\end{figure}

\begin{figure}
  \includegraphics[width=16 cm]{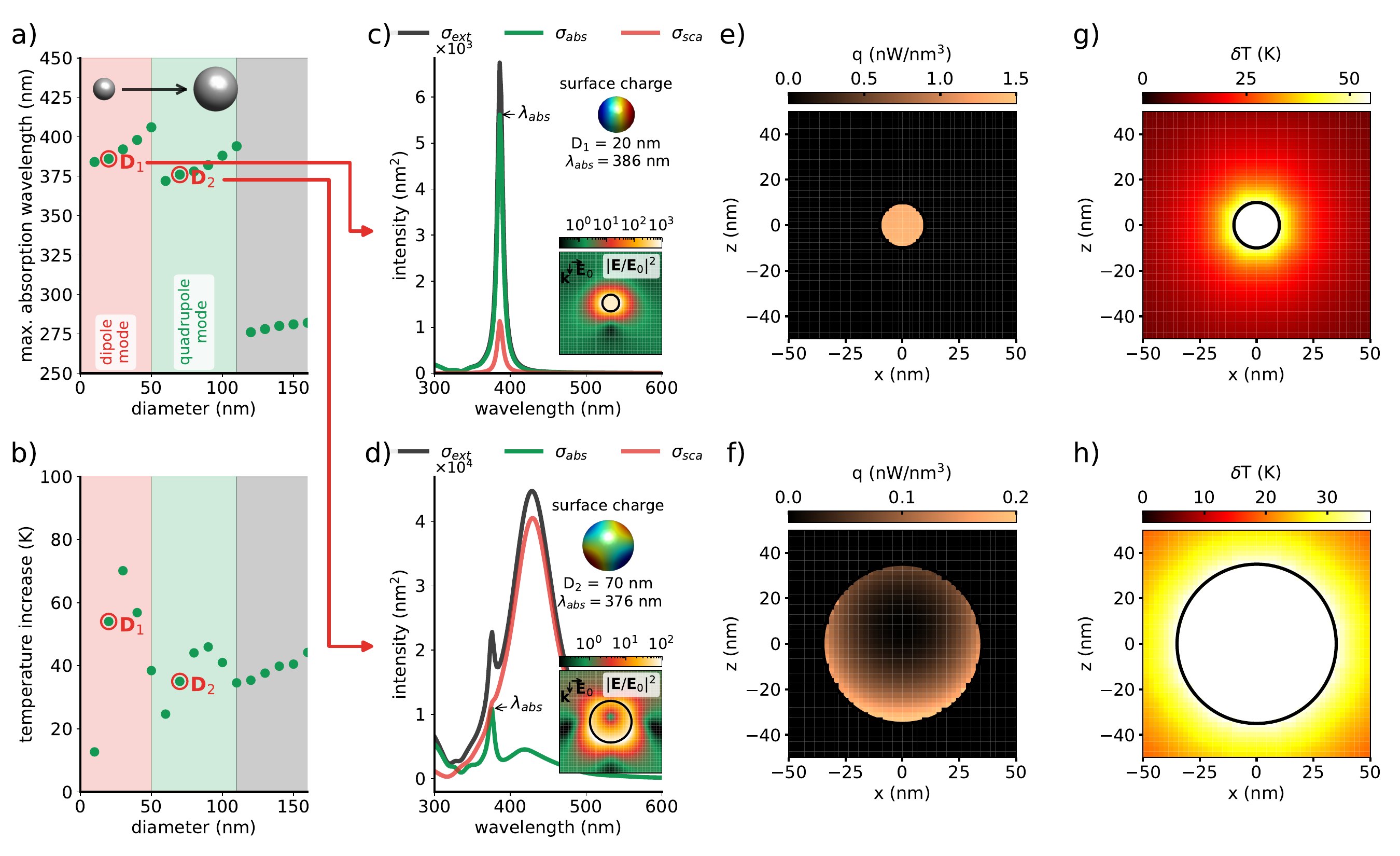}
  \caption{Thermoplasmonic response of a single Ag nanoparticle: diameter dependence. (a) Wavelength of the maximum absorption cross-section peak ($\lambda_{abs}^{max}$) as a function of nanoparticle diameter. (b) Temperature increase ($\delta$T) at $\lambda_{abs}^{max}$ as a function of nanoparticle diameter, excitation irradiance was $10^9$ W/m$^2$. (c, d) Spectra of the extinction, absorption, and scattering cross-sections for Ag nanoparticles with 20 nm (c) and 70 nm (d) diameters. Insets show the corresponding surface charge distribution and the electric field map at $\lambda_{abs}^{max}$. (e, f) Heat power density (q) maps for the AgNPs with 20 nm (e) and 70 nm (f) diameters. (g, h) Temperature increase for the Ag nanoparticles with 20 nm (g) and 70 nm (h) diameters.}
  \label{SIFig:AgNP_single_res}
\end{figure}

\begin{figure}
  \includegraphics[width=8 cm]{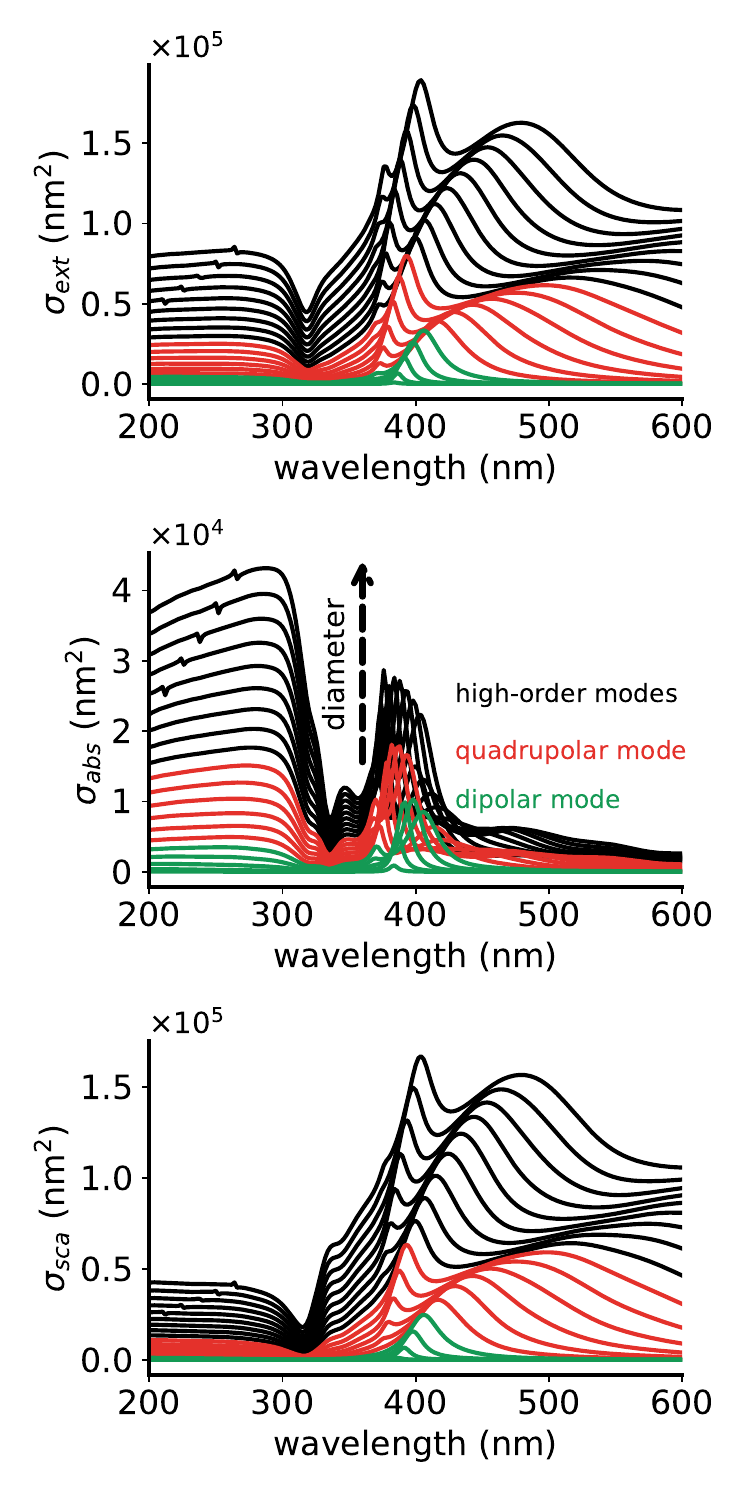}
  \caption{Extinction, absorption, and scattering cross-section spectra of a single silver nanoparticle with diameter varied from 10 to 200 nm. The color of the spectra (green, red, black) denotes the dominant contributing plasmon resonance: dipole, quadrupole, and higher-order modes, respectively.}
  \label{SIfig:CS_single_AgNP}
\end{figure}

\begin{figure}
  \includegraphics[width=16 cm]{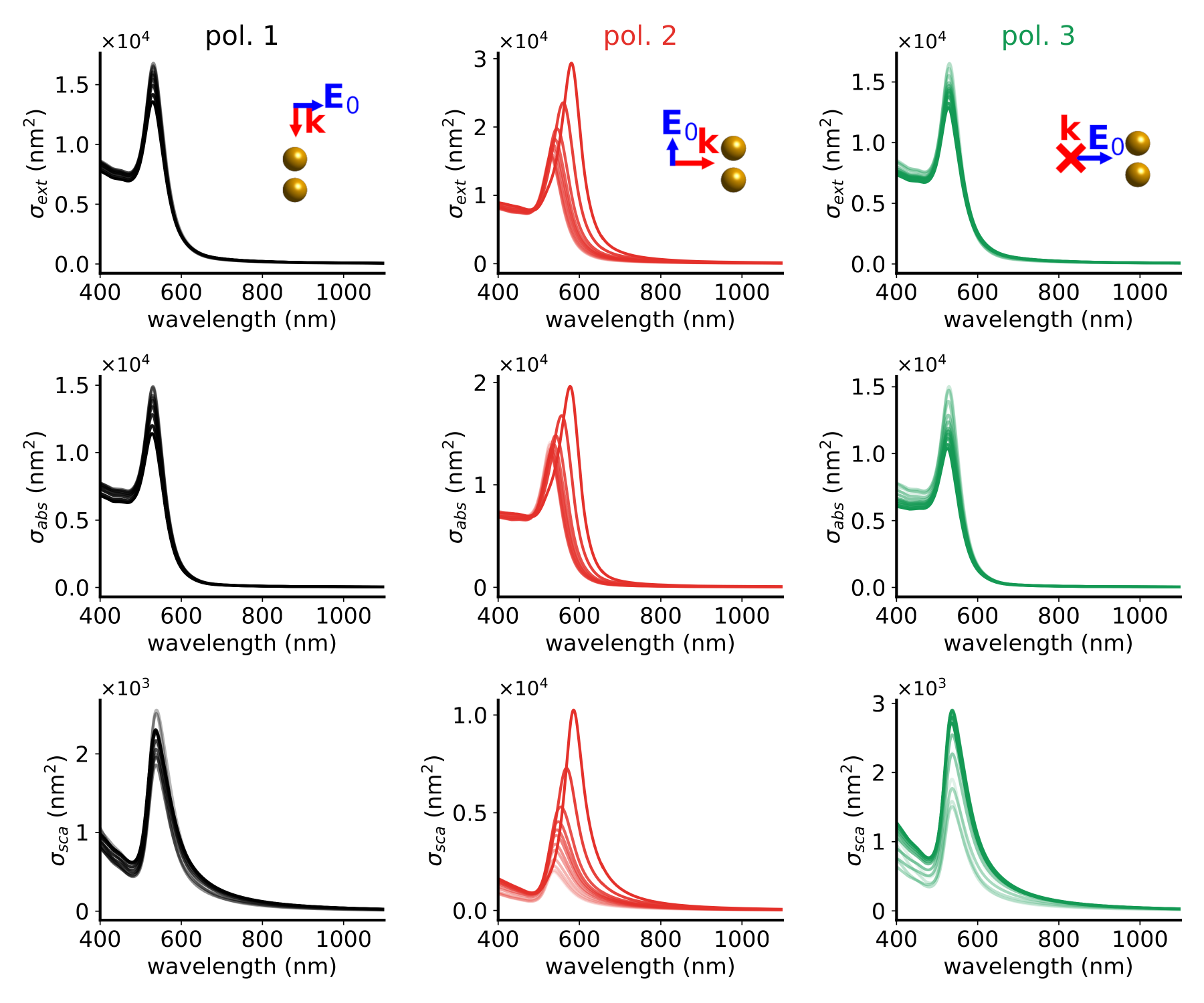}
  \caption{Extinction, absorption, and scattering cross-section spectra of a gold nanoparticle dimer under three orthogonal incident polarizations: transverse to the dimer axis (pol. 1, black), longitudinal along the dimer axis (pol. 2, red), and at 90$^0$ orientation to the axis (pol. 3, green). Each nanoparticle has a diameter of 50 nm, with an interparticle gap varied from 1 to 500 nm (spectrum color saturation from darker to lighter).}
  \label{SIfig:CS_dimer_AuNP}
\end{figure}

\begin{figure}
  \includegraphics[width=16 cm]{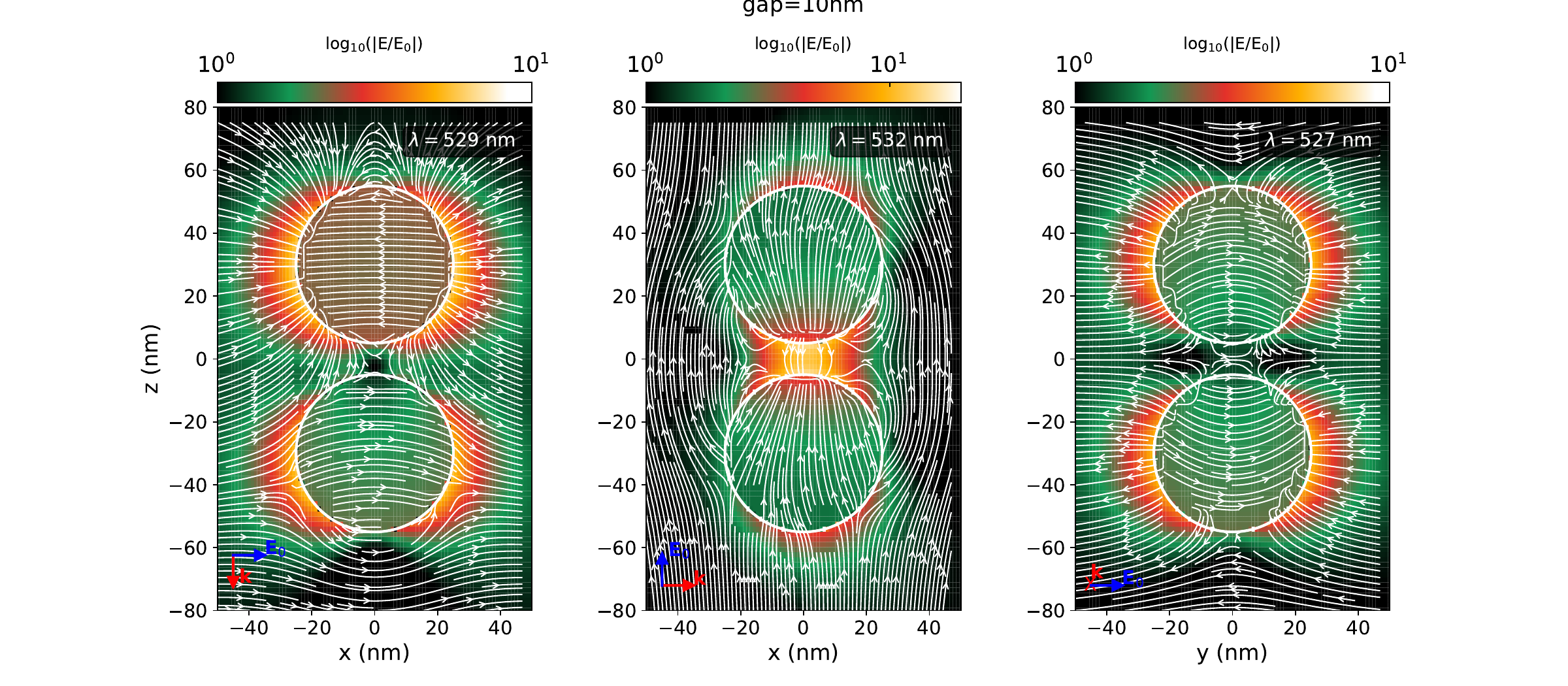}
  \caption{Electric field enhancement ($\text{log}|\textbf{E}/\textbf{E}_0|^2$) and electric field lines in the plane of excitation for a dimer  with a 10-nm gap under three incident polarizations. Each nanoparticle has a diameter of 50 nm. White circles in indicate nanoparticle boundaries, while red and blue arrows show the incident wave vector ($\textbf{k}$) and incident electric field ($\textbf{E}_0$) directions, respectively.}
  \label{SIFig:AuNP_Efield_dimer_gsmall_res}
\end{figure}

\begin{figure}
  \includegraphics[width=16 cm]{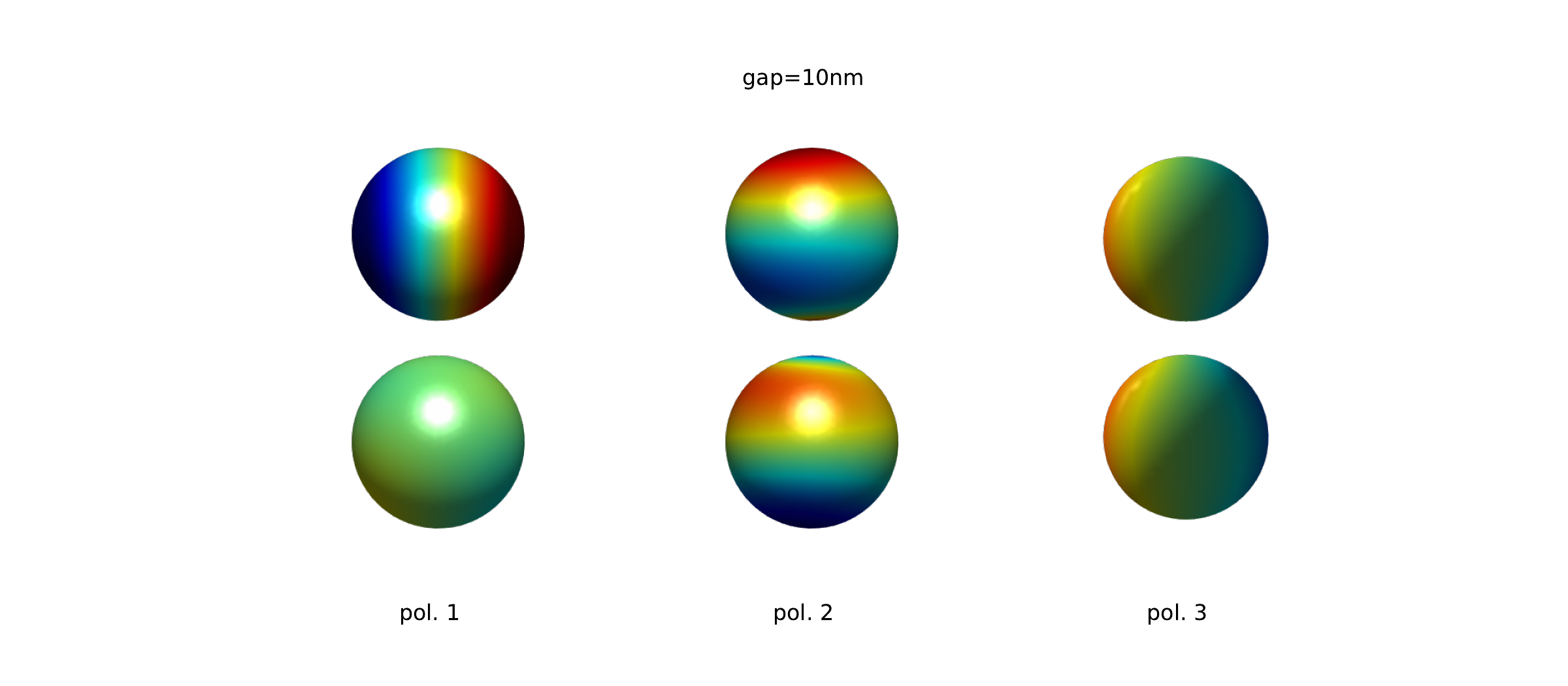}
  \caption{Surface charge density distribution for a dimer  with a 10-nm gap under three incident polarizations. Each nanoparticle has a diameter of 50 nm. For polarization 3, the dimer is rotated $90^0$ around the z-axis to visualize the charge distribution.}
  \label{SIFig:AuNP_sigma_dimer_gsmall_res}
\end{figure}

\begin{figure}
  \includegraphics[width=16 cm]{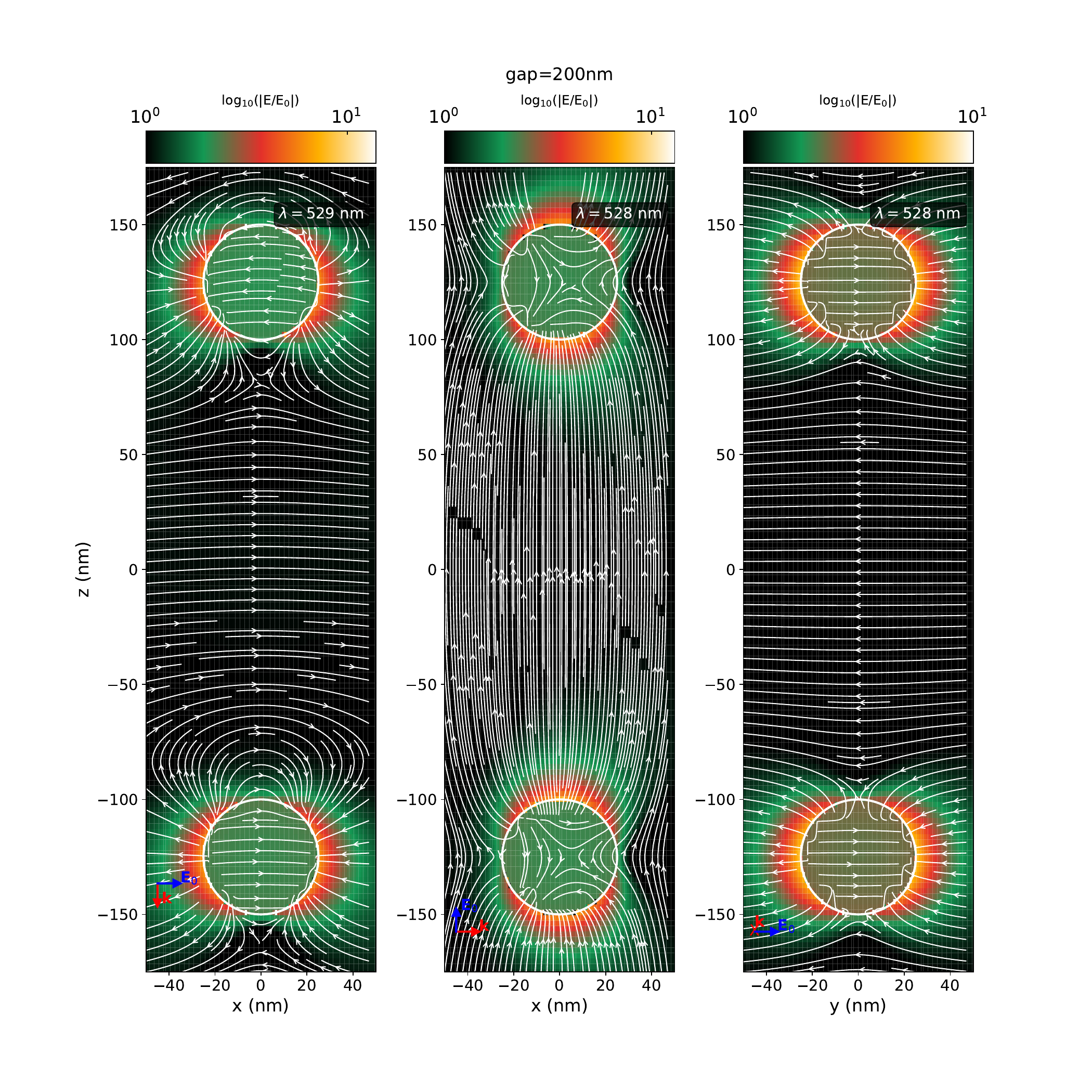}
  \caption{Electric field enhancement ($\text{log}|\textbf{E}/\textbf{E}_0|^2$) and electric field lines in the plane of excitation for a dimer  with a 200-nm gap under three incident polarizations. Each nanoparticle has a diameter of 50 nm. White circles indicate nanoparticle boundaries, while red and blue arrows show the incident wave vector ($\textbf{k}$) and incident electric field ($\textbf{E}_0$) directions, respectively.}
  \label{SIFig:AuNP_Efield_dimer_gbig_res}
\end{figure}

\begin{figure}
  \includegraphics[width=16 cm]{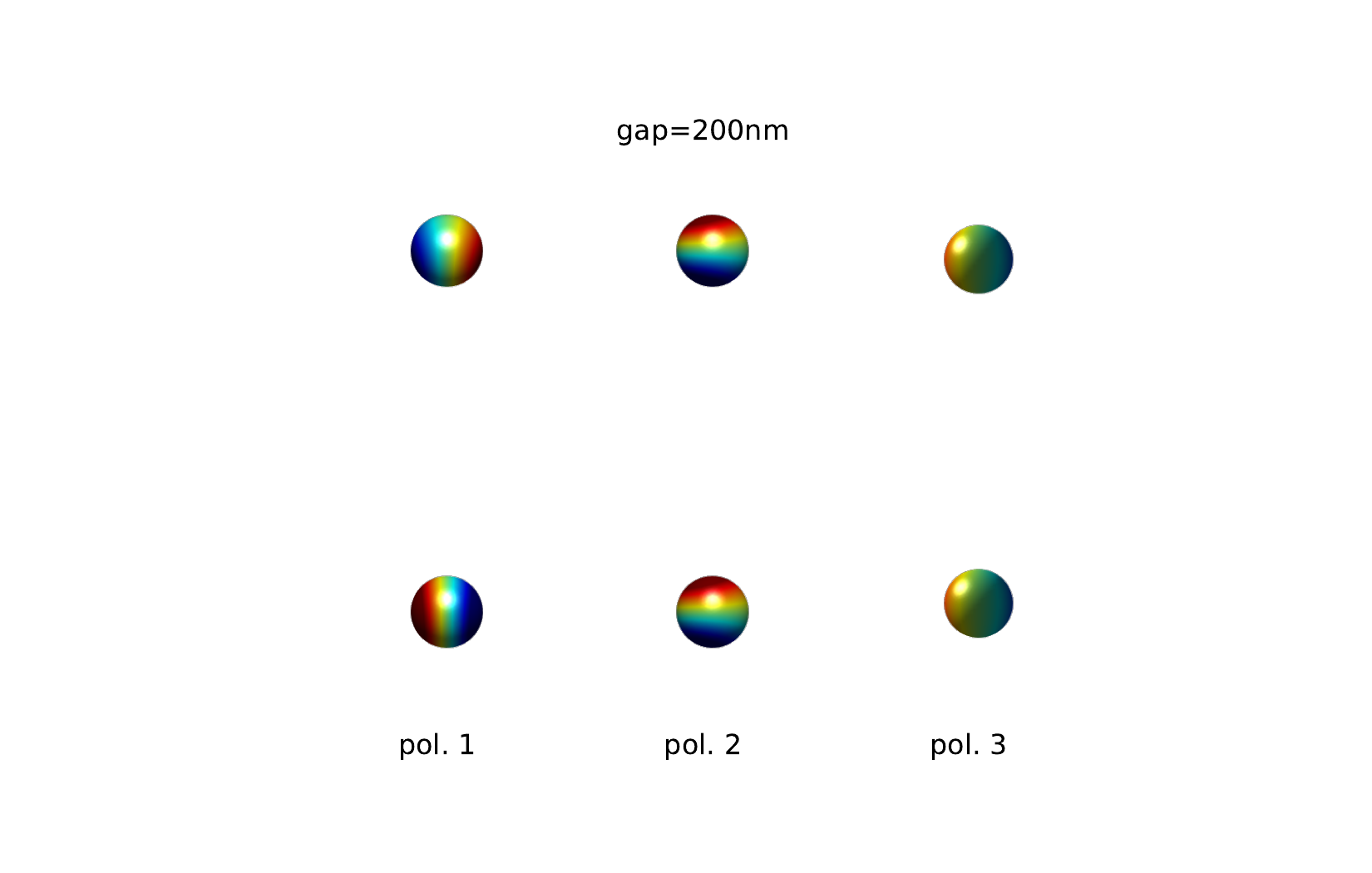}
  \caption{Surface charge density distribution for a dimer  with a 200-nm gap under three incident polarizations. Each nanoparticle has a diameter of 50 nm. For polarization 3, the dimer is rotated $90^0$ around the z-axis to visualize the
charge distribution.}
  \label{SIFig:AuNP_sigma_dimer_gbig_res}
\end{figure}

\begin{figure}
  \includegraphics[width=12 cm]{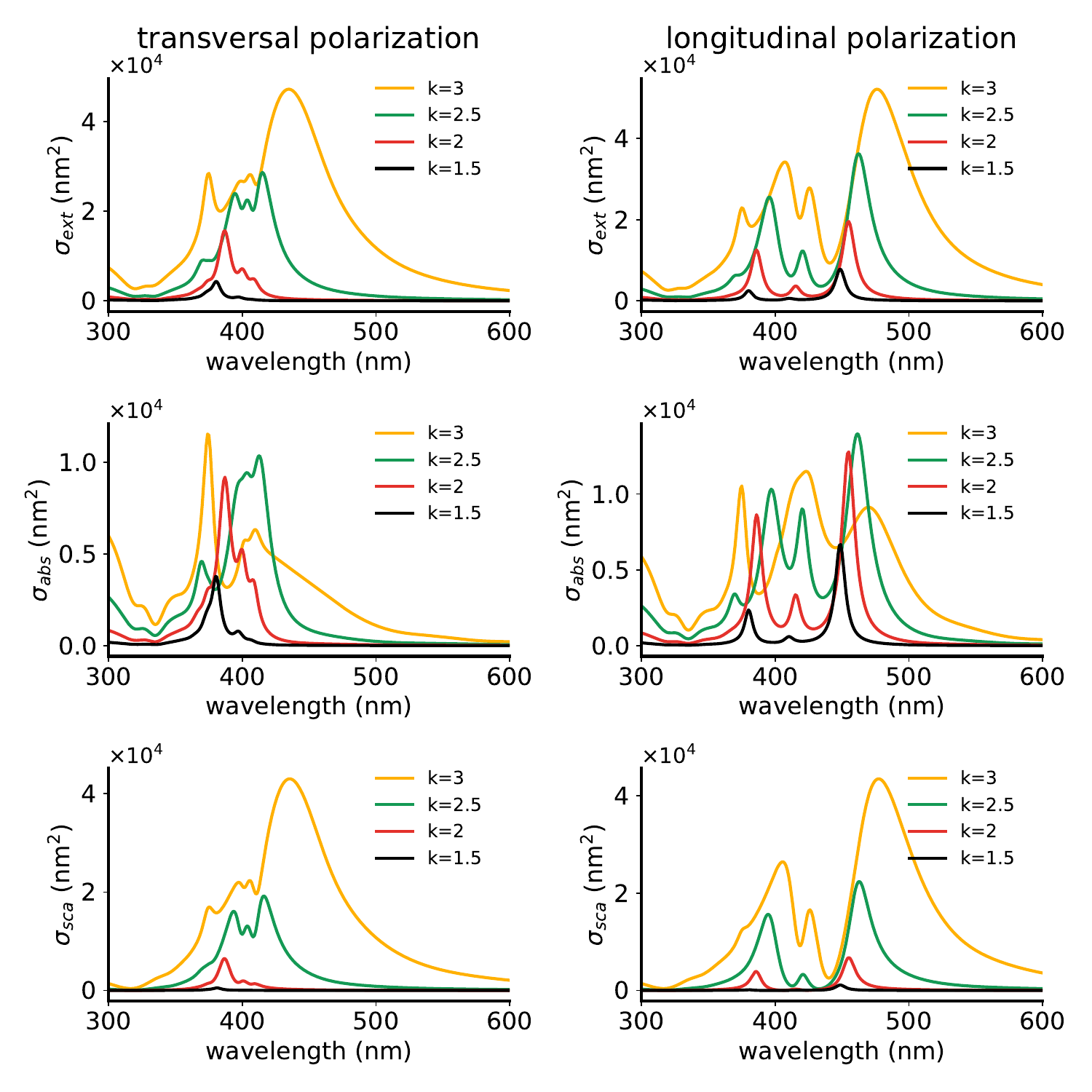}
  \caption{Optical cross-section spectra of a silver nanolens trimer. The structure consists of three nanoparticles with diameters and gaps scaled by a geometrical self-factor $\kappa$ (1.5, 2, 2.5, 3), see scheme in Figure 3b in the main text. Spectra are shown for two incident polarizations: transverse (solid lines) and longitudinal (dashed lines) to the trimer axis.}
  \label{SIfig:CS_nanolens_AgNP}
\end{figure}

\begin{figure}
  \includegraphics[width=16 cm]{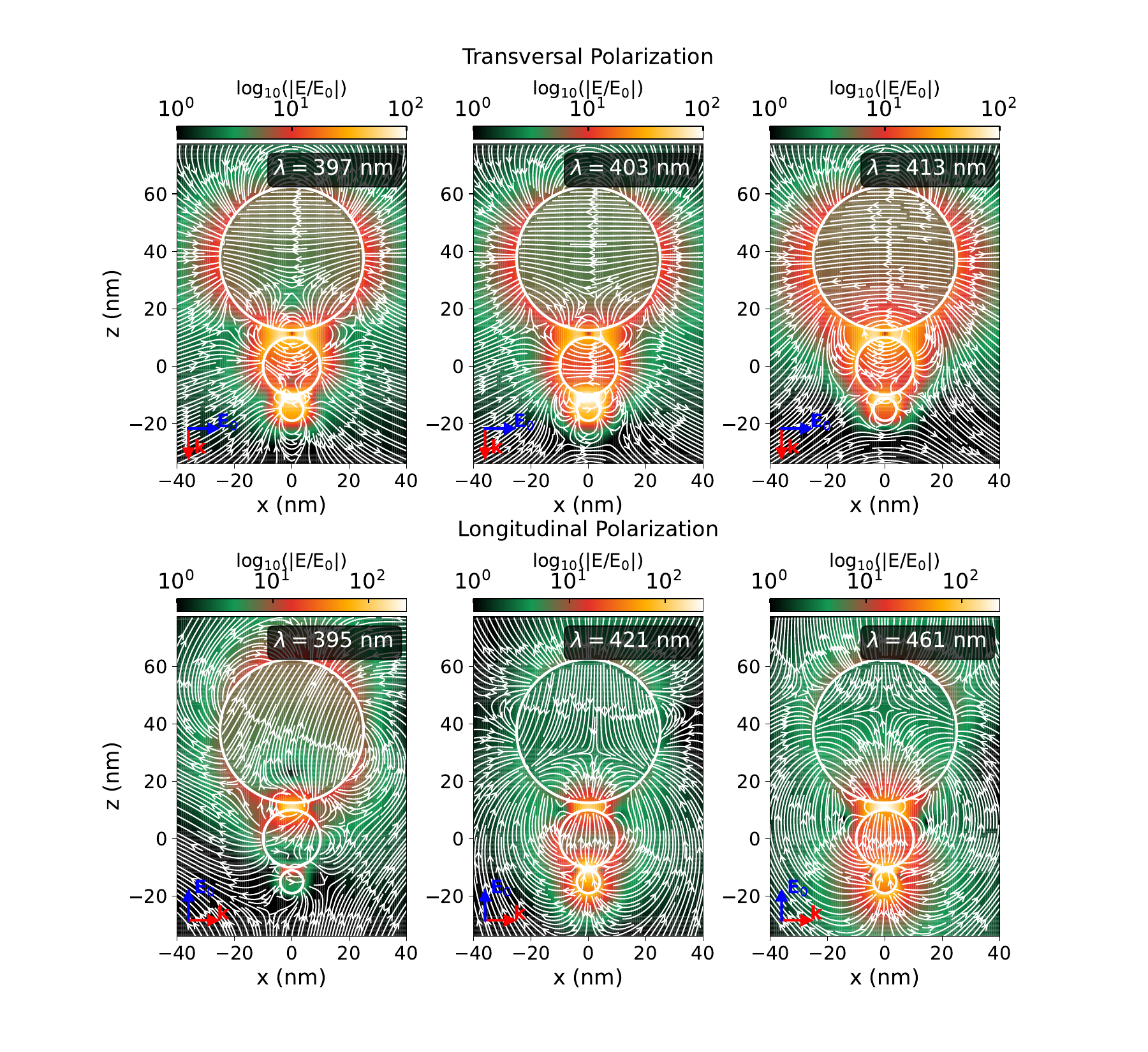}
  \caption{Electric field enhancement and electric field lines for a silver nanolens trimer with $\kappa=2.5$. The map shows $\text{log}|\textbf{E}/\textbf{E}_0|^2$ in the xz-plane at the dominant absorption peaks for transverse and longitudinal incident polarizations. White circles indicate nanoparticle boundaries, while arrows show the incident wave vector ($\textbf{k}$) and incident electric field ($\textbf{E}_0$) directions.}
  \label{SIFig:AgNP_Efield_nanolens_res}
\end{figure}

\begin{figure}
  \includegraphics[width=16 cm]{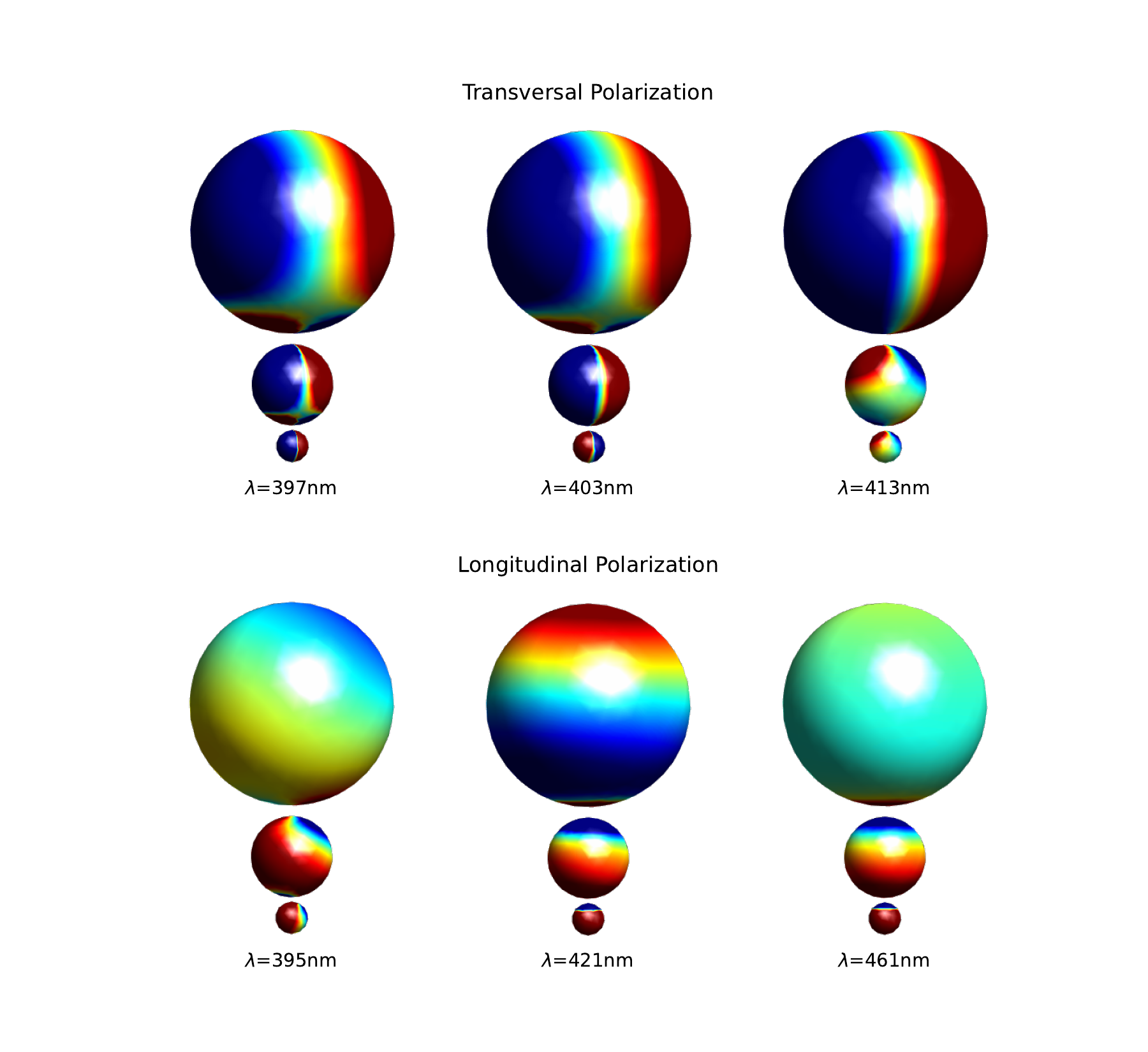}
  \caption{Surface charge distribution of the silver nanolens trimer with $\kappa=2.5$ at the dominant absorption peaks. The top and bottom rows correspond to transverse and longitudinal incident polarizations, respectively. Each column shows the charge distribution for a specific resonant mode, ordered by increasing wavelength. Red and blue colors indicate positive and negative induced charge densities.}
  \label{SIFig:AgNP_sigma_nanolens_res}
\end{figure}

\clearpage

\bibliography{References}